\documentclass[aps,prb,twocolumn,longbibliography,notitlepage,superscriptaddress,nofootinbib]{revtex4-1}

\bibpunct{[}{]}{;}{n}{}{}
\usepackage[caption=false]{subfig}

\usepackage{./MyStyle/myStyle}
\def\eps{\epsilon}

\usepackage{ gensymb }
\usepackage{xfrac}
\usepackage{slashed}
\usepackage{framed}
\usepackage{hhline}
\usepackage{footnotebackref}

\usepackage{tcolorbox}
\usepackage{cleveref}
\usepackage{graphicx}

\def\moire{\text{moir\'e}}
\def\eph{e_{\text{ph}}}

\def\Nel{N_{\text{el}}}
\graphicspath{ {./Figures/} }
\def\Enot{E_\circ}
\def\Bp{\bar{\Bmc}}

\def\mst{m^\star}
\def\Kmc{\mathcal{K}}

\def\ee{\mathrm{e}}
\def\ii{\mathrm{i}}

\def\rs{r_\mathrm{s}}
\DeclareMathOperator\Erfc{Erfc}
\def\OO{\mathrm{O}}

\DeclareMathOperator{\sinhc}{sinhc}
\def\eff{\text{eff}}
\def\BZWC{\mmrm{BZ}}
\def\WC{\text{WC}}
\def\WBZ{\nu'}
\def\Wwc{\W}
\def\Qwc{\q_{\text{WC}}}

\def\kWC{k_{\circ}}

\begin{document}

\title{
Pseudo-spin order of Wigner crystals in multi-valley electron gases
}
\author{Vladimir Calvera} 
\email{fvcalvera@stanford.edu}
\author{ Steven A. Kivelson }
\email{kivelson@stanford.edu}
\affiliation{Department of Physics, Stanford University, Stanford, CA 94305, USA}
\author{Erez Berg}
\email{erez.berg@weizmann.ac.il.}
\affiliation{Department of Condensed Matter Physics, Weizmann Institute of Science, Rehovot 76100, Israel}
\begin{abstract}
    We study multi-valley electron gases in the low density ($r_s \gg 1$) limit. Here the ground-state is always a Wigner crystal (WC), with additional pseudo-spin order where the pseudo-spins are related to valley occupancies. Depending on the symmetries of the host semiconductor and the values of the parameters such as the anisotropy of the effective mass tensors, we find  a striped or chiral pseudo-spin antiferromagnet, or a time-reversal symmetry breaking orbital loop-current ordered pseudo-spin ferromagnet. Our theory applies to the recently-discovered WC states in AlAs and in mono and bilayer transition metal dichalcogenides. We identify a set of interesting electronic liquid crystalline phases that could arise by continuous quantum melting of such WCs.
\end{abstract}

\maketitle

\section{Introduction}

The study of the two dimensional electron gas (2DEG), both experimentally and theoretically, has been extraordinarily fruitful \cite{RevModPhys.54.437}.  From the theoretical perspective, it is one of the most basic problems in correlated electron physics, while because it can be realized and probed in semiconductor devices, it is of direct experimental significance.  In its simplest realization, the electrons are associated with a unique minimum (``valley'') in the host semiconductor's band dispersion, and can be treated (in the context of the effective mass approximation) as formally analogous to electrons in free space.  

However, in many experimentally relevant cases, the host semiconductor has more than one relevant valley \cite{shayegan2006two}.  This results in a more structured version of the 2DEG in which the electrons carry a pseudo-spin index corresponding to the distinct valleys. This opens the possibility of a host of new collective states.  

In the present paper, we analyze various natural multi-valley versions of the 2DEG in the low density (large $r_s$) limit.  Here, the Coulomb interactions dominate the kinetic energy, and so for precisely the same reasons as in the simple 2DEG, the system forms an insulating, Wigner crystalline (WC) state \cite{PhysRev.46.1002}.  However, when the electrons in question have a pseudo-spin degree of freedom, the nature of the pseudo-spin order in the WC remains to be resolved. Here, we address this issue for various natural valley structures (i.e. for different sorts of pseudo-spin symmetries).  We show that the pseudo-spin order can be determined in an asymptotically controlled expansion in $\sqrt{1/r_s}$.  (By contrast, 
the issue of spin order in the WC cannot be addressed in any order in $\sqrt{1/r_s}$, as it involves tunneling processes of order $\exp[ - \alpha \sqrt{r_s}]$.) {Analogous phenomena have been discussed in the context of 2DEGs with Rashba spin-orbit coupling, that can be viewed as having a continuous, rather than discrete, pseudo-spin degree of freedom~\cite{Berg2012Rashba2DEGcrystal,PhysRevB.89.155103}}.

As addenda, we consider several extensions of these results: 1) We comment on the application of these ideas to moir\'{e} systems, in which commensurability effects between the WC and the moir\'{e} lattices can be tuned.  2) We briefly consider the simplest 3d version of the  pseudo-spin ordering problem.  3) We consider some interesting possible multi-step quantum or thermal melting  processes of a pseudo-spin ordered 2D WC that  lead to  electronic liquid crystalline states with various patterns of vestigial order. 
\begin{figure}[h]
     \includegraphics[width=0.9\columnwidth]{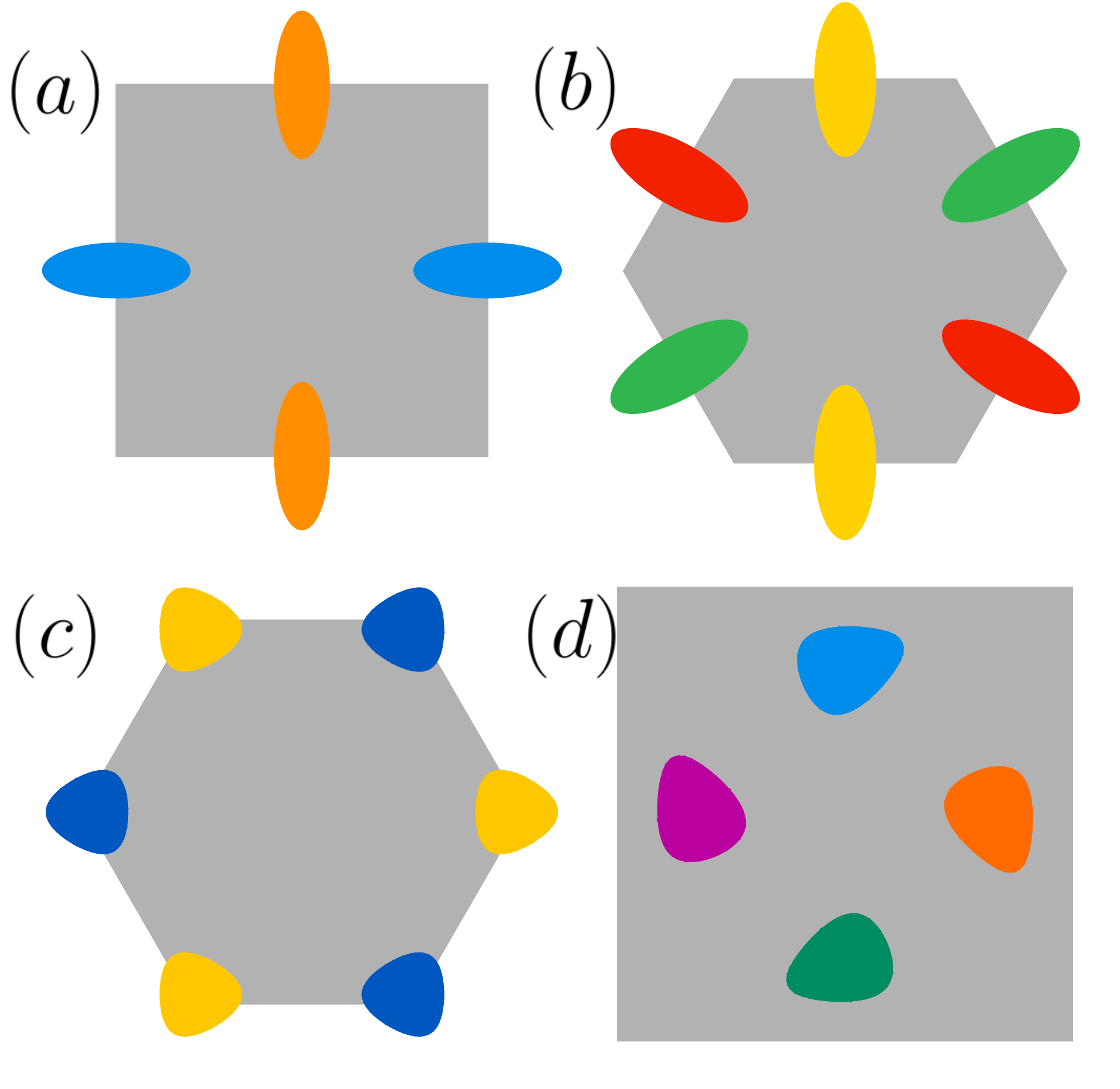}
    \caption{Schematic representation of valleys on the Brillouin zone of 2d crystals. The distinct colors indicate distinct valleys (pseudo-spin polarizations) and the shapes are representative of the Fermi surface at some fixed, small electron density. Cases (a) and (b) correspond to valleys with anisotropic masses tensors (Sec.~\ref{sec:AnisotropicMassTensor}). Case $(c)$ correspond to a pair of valleys related by inversion symmetry and cubic corrections to the dispersion relations (Sec.~\ref{sec:InversionPairs}). Case $(d)$ is a more complicated example that could be realized on a system with $C_4$ rotation but no reflection plane. 
    \label{fig:valleys} The same color scheme will be used in subsequent figures. 
    }
\end{figure}

The paper is organized as follows. In Sec.~\ref{sec:background} we introduce some notation and review the approximations used in this work. Sec.~\ref{sec:AnisotropicMassTensor} considers the effect of mass anisotropy on the Wigner crystal and how the pseudo-spin degeneracy is broken when there are multiple valleys related by rotations. Sec.~\ref{sec:InversionPairs} treats possible ways to lift the degeneracy between pairs of valleys related by inversion symmetry. In Sec.~\ref{sec:Extensions}, we consider extensions of our considerations to external moir\'{e} potentials and to Wigner crystals in three dimensions. Then in Sec.~\ref{sec:Melting} we entertain various scenarios for melting of our Wigner crystal. 
The bulk of the paper is relatively technical, so in Sec.~\ref{sec:Summary} we summarize the principle findings.

\section{Background}\label{sec:background}

\subsection{Multi-valley semiconductors}
Consider a semiconductor whose conduction band has minima at 
$n_v$ distinct, symmetry-related momenta $\{\Qbs_{1},\dots,\Qbs_{n_v}\}$ 
in the Brillouin zone. When we slightly dope the system we can use the effective mass approximation. 
Here we treat the conduction electrons with crystal momenta near $\Qbs_{\alpha}$ ($\kbs\approx\Qbs_\alpha$) as one flavor of a 2-dimensional electron gas with dispersion relation 
\begin{equation}\label{eq:KineticEnergyEffectiveMass}
T_\alpha({\bf p})= \sum_{a,b}\frac{1}{2}W_{ab}^{(\alpha)}p_{a}p_b
\end{equation}
where in terms of $\e_{\kbs}$, the electron dispersion relation,  the effective inverse mass tensor is defined as
\begin{equation}
    W^{(\alpha)}_{ab}=\pdv{\e_{\kbs}}{k_a}{k_b}\eval_{\kbs=\Qbs_\alpha} .
\end{equation}
We will refer to the electrons with lattice momentum near $\Qbf_\alpha$ to have \textit{valley flavour} $\Qbf_\alpha$ or simply $\alpha$. {Here and henceforth we work in units where $\hbar =1$.} 

The point-group (PG) symmetry of the semiconductor imposes the following constraints: If $\gbf\in PG$ acts as $\rbf \rightarrow R_\gbf \rbf$ in real space, then $\gbf$ must permute the valleys, $R_\gbf Q_\alpha =Q_\alpha'$ (modulo reciprocal lattice vectors), and the mass tensors are related by
\begin{equation}
    W^{(\alpha)}=R_{\gbf}\cdot W^{(\alpha)}\cdot R_{\gbf}^{\top}.
\end{equation}
Manifestly, the inverse mass tensors $W^{(i)}$ need not 
have the full PG symmetry of the lattice, but only the  
subgroup 
that 
leaves ${\bf Q}_{\alpha}$ invariant.

In this paper we consider 
electrons interacting via long-range Coulomb interactions that do not depend on the valley flavor \footnote{We ignore short range interactions that depend on the valley flavour because these corrections are suppressed in the low density limit with respect to the contributions we take into account.   } so
\begin{equation}
    V(\Rbs-\Rbs')= \frac{e^2}{4\pi \eps}\frac{1}{\norm{\Rbs-\Rbs'}}.
\end{equation}
%
%
The true symmetries of the system are 
those of the lattice Hamiltonian which has a space group (discrete translations combined with the PG)
and internal symmetries such as time reversal symmetry and spin rotation. Once we 
make the effective mass approximation, there is an emergent continuous translation symmetry $(\RR^2)$ and particle number conservation for each valley ($\U(1)_{(i)}$). Note that the diagonal subgroup of $\prod_{i=1}^{n_v}\U(1)_{i} $ should be identified with physical particle number conservation, and is thus an exact symmetry. 

In the familiar case of a single valley at a high symmetry point in the Brillouin zone, the emergent symmetry includes the full $O(2)$ rotational symmetry of Euclidean space, but in most cases this does not happen. Indeed, the individual $W^{(i)}$'s do not  have the full PG symmetry; typically, PG operations operate both on the spatial indices $a$ and $b$, as well as involving permutation of the flavor indices, $i$. (In some cases the emergent symmetry is even larger, e.g. there is a $\U(2)$ symmetry when there are two different valleys related by inversion symmetry, but we will not consider this cases here.) For simplicity, we will neglect the possibility of non-symmorphic symmetries, so the effective mass Hamiltonian has the symmetry group
\[
G_{EM}=(\RR^2\times \prod_{i=1}^{n_v} \U(1)_{(i)})\rtimes PG  \times G_{\text{int}}
\]
where $PG$ acts as $O(2)$ on $\RR^2$ and by permutation on the product of $\U(1)$'s. Here $G_{\text{int}}$ are the internal symmetries with total particle conservation omitted, which can include 
spin-rotation and time-reversal symmetries.

\subsection{Low-density limit}
\def\aeff{a_{\eff}}
\def\rs{r_{\text{s}}}
\def\nel{n_{\text{el}}}
The discussion so far holds for 
electron densities low enough compared to atomic scales, 
thus justifying the effective mass approximation. There is 
another density scale that 
determines the balance between the
kinetic energy and the Coulomb interactions, 
conventionally defined as $1/\pi \aeff^2 $ where $\aeff = \frac{4\pi \e_0 \hbar^2}{e^2 m_{\eff}}$ is the effective Bohr radius of the system in which $m_{eff}$ is the square root of the determinant of the effective mass tensor.
We define the dimensionless quantity $\rs = (\pi \nel \aeff^2)^{-\sfrac{1}{2}}$ as usual. We focus on the $\rs\gg 1$ regime, i.e. $\nel \ll [\p \aeff^2]^{-1}$. 

As in the isotropic 2DEG, the kinetic energy per electron scales as $1/\rs^2$ while the Coulomb energy per electron scales as $1/\rs$ but the prefactors depend on the various mass anisotropies. 
For large values of $\rs$ the Coulomb energy dominates forcing the electrons to form a triangular lattice \cite{Bonsall1977WignerCrystal2D}. 
For later convenience we define $\nu$ as the area per electron, which for the triangular lattice is $\nu=\frac{\sqrt{3}} {2}a^2$,  {in which, by appropriate choice of units, we will set the WC lattice constant $a=1$}. 

At this point, there is an extensive degeneracy associated with the pseudo-spin orientation at every lattice site, the resolution of which is the primary goal of the present work.  In the absence of spin-orbit coupling, there is also an extensive degeneracy associated with spin orientations, which we will ignore for the bulk of this paper. 
{\footnote{The extensive spin-related degeneracies that arise in the absence of exchange interactions persist even in the presence of spin-orbit coupling so long as  inversion and/or time-reversal symmetry are preserved;  where these symmetries are broken, spin-valley locking can lift this degeneracy even in the absence of exchange interactions.}}
Finally, there are the usual global degeneracies associated with the broken translation and rotation symmetries in the WC. 
The former of these is exact within effective mass approximation, but (as we will discuss in the context of moir\'{e} systems) is really only a discrete symmetry when commensurability effects with the underlying lattice are included.  However, even within the effective mass approximation, the rotational symmetry of the classical ($r_s \to \infty$) problem will be resolved to the discrete PG symmetry, when pseudo-spin ordering is considered; it is thus important to define the orientation of the WC in terms of an angle defined relative to the symmetry axes of the host crystal ($\Qwc$). 

The following are the principle features of the band dispersion that could control the pattern of symmetry breaking in the large $r_s$ limit according to the order in $1/r_s$ at which they arise:
\begin{enumerate}
    \item Valleys with different mass tensors (order - $1/\rs^{3/2}$);
    \item {Trigonal warping corrections to the dispersion relations} (order - $1/\rs^{3}$);
    \item Berry curvature effects - although they are unimportant for the examples we have considered (order - $1/\rs^{3}$ or $1/\rs^{9/2}$).
\end{enumerate}


We will proceed as follows: First, we consider the effect of feature 1 by studying the phonon zero point energy { as a function of the mass anisotropy, $\lambda \equiv [{m_L-m_T}] /  [{m_L+m_T}]$, for $0 < \lambda \leq 1$ ($m_{L}$ and $m_{T}$ are the eigenvalues of the mass tensor. When the valley lies on a reflection-symmetric axis of the BZ, we can identify $m_L$ and $m_T$ as the longitudinal  and transversal masses, respectively.).  For small $\lambda$, we obtain an effective statistical mechanics model from which the optimal pattern of pseudo-spin ordering can be derived.  For more general $\lambda$, we evaluate the energy of this and other possible small period patterns of pseudo-spin order to argue that the same pattern of pseudo-spin order is optimal over the entire range of $\lambda$.
}


Next, we add the corrections to the dispersion relations (features 2 and 3) as  small perturbations (in $1/\rs$) to the previous states. We again obtain an effective statistical model for the pseudo-spins that determines 
more subtle features of the ground-state order. 



\subsection{Effective mass and harmonic approximations}
\label{sec:EffectiveMassHarmonicApprox}

The effective Hamiltonian has the form 
\begin{align}
    H = \sum_j T_{\alpha_j}(\vec p_j) + \sum_{i>j}V(\vec r_i -\vec r_j- \vec R_{i} + \vec R_j),
\end{align}
where $j$ labels an electron in valley $\alpha_j$ with momentum $\vec p_j$ at position $\vec r_j + \vec R_j$ ($\vec R_j$ is the classical WC ground state position). 

To quadratic order in these displacements, we 
consider dimensionless position/momentum variables as $P_{ia} = \frac{p_{ia}}{\sqrt{\Enot \mst}} $ and $Q_{ia}=\sqrt{\Enot \mst} q_{ia}$ where $x_{ia}$ is the electron displacement in the $a$ direction away from 
the classical equilibrium position $\vec{R}_i$ and $p_{ia}$ is the conjugate momentum. We use 
\begin{equation}
    \mst={\frac{2m_Lm_T}{m_L+m_T}};\Enot =\sqrt{\frac{e^2 \hbar^2}{4\pi \eps a^3 \mst }}   \propto \rs^{-\sfrac{3}{2}}
\end{equation}
for mass and energy scales, respectively. The effective dimensionless Hamiltonian is
\begin{equation}\label{eq:HamiltonianPhonon}
    h_{\text{eff}}=\frac{H_{\text{eff}}}{ \Enot} = \sum_{i} \frac{1}{2}P_{i}\overline{W}_{i}P_i +\sum_{i, j} \frac{1}{2} Q_iK_{ij}Q_j
\end{equation}
where $\overline{W}_i$ is $W^{(\a_i)}/\mst$ and $K_{ij}$ is the dipole-dipole coupling matrix given by 
\begin{equation}
    \begin{split}
    (K_{ij})_{ab} &= \frac{1 }{R_{ij}^3}\left(\d_{ab} - 3 \frac{(R_{ij})_a(R_{ij})_b}{R_{ij}^2}\right); i\neq j\\ 
    (K_{ii})_{ab}&= \gamma \d_{ab};\quad \gamma =\frac{1}{2} \sum_{i\neq 0} \frac{1}{R_{i0}^3}. 
    \end{split}
\end{equation}
Where $\g \approx 5.5171$ for the triangular lattice. 
{
It is important to recognize that the effective mass tensor in Eq. \ref{eq:HamiltonianPhonon} depends implicitly on the pseudo-spin ordering.  Because the effective Hamiltonian is quadratic, it can still be diagonalized in terms of normal modes (phonons) with energies $\hbar \omega_\mu$.  Thus, the pseudo-spin-configuration dependent zero point energy of the phonon per electron can be expressed as
\begin{align}\label{eq:e_eff}
   e_{\text{ph}}\equiv\frac{E_{\text{eff}}}{\Nel \Enot} = \frac 1 {2\Nel} \sum_{\mu} \omega_{\mu}
\end{align}
where $E_{\text{eff}}$ is the phonon zero-point motion correction to the ground-state energy per electron.  (Similar expressions can be obtained for the configuration dependent free energy, at low enough temperatures that the harmonic approximation is reliable.)  The preferred pseudo-spin order is the pattern which minimizes $E_{\text{eff}}$. {(This is, in essence, a form of ``order by disorder.'')}}

{
Note that when the pseudo-spin pattern is periodic in space, we can label the normal modes by a crystal momentum and a band index. The crystal momenta live in a new Brillouin zone (BZ$'$) that is obtained by folding the original BZ of WC. The band index take $2B$ values, where $B$ is the number of lattice sites in the unit cell of the pattern. 
}

\section{Anisotropic mass tensors}\label{sec:AnisotropicMassTensor}

In this section, we calculate the effect on Wigner crystallization when the mass tensors are anisotropic and the valleys are related by rotations. {Recall that we measure the anisotropy using the parameter
\begin{equation}
    \lambda = \frac{m_L-m_T}{m_L+m_T}
\end{equation}
where $m_L>m_T$ are the two eigenvalues of the effective mass tensor\footnote{We use the subscripts $L$ and $T$ because in the examples we have in mind (Figs.~\ref{fig:valleys}{\blue {(a)}} and \ref{fig:valleys}{\blue{(b)}}) the two masses correspond to the longitudinal and transversal masses. }.
} The outline of this section is as follows. First, we consider the small $\lambda$ limit and derive an effective statistical model. Next, we use a variational approach to study other values of $\lambda$. Finally, we apply the results to cases relevant to experiments where there are $n_{v}=1,2$ or $3$ valleys.

\subsection{Small mass anisotropy: perturbative approach}\label{sec:SmallMassAniPerturbative}

We calculate the leading in $\lambda \ll 1$ behavior of the system by mapping the problem to a classical clock model on the triangular lattice with long-range angle-dependent interactions. See App.~\ref{app:MapToClockModel} for details.

{ The phonon zero-point energy in Eq.~\ref{eq:e_eff} can be expanded in powers of $\lambda$ as $e_{\text{ph}}=\sum_{n=0}^{\infty} e_{n}\lambda^n$ with coefficients that depend on the pattern of orbital ordering.  This, in turn, is most efficiently represented by identifying a  phase variable, $\theta_j$ on each site, that indicates the orientation of the preferred axis of the effective mass tensor, and which can take on one of $n_v$ discrete values, equally spaced in units of $2\pi/n_v$ starting from $\theta_{\WC}$ which is the angle between one of axis of the WC and the host semiconductor. The leading correction that depends on the valley pattern is $\lambda^2 e_2$, where
\begin{equation}\label{eq:XYenergy}
   e_2 = -\frac{1}{\Nel}\sum_{\kbs} \Phi^{\dag}_{2;\kbs}\tilde{J}(\kbs)\Phi_{2;\kbs}^{},
\end{equation}
$\Phi_{2;\kbs}$ is the Fourier transform of $[\cos(2\q_i),\sin(2\q_i)]^{\top}$ and $\tilde{J}(\kbs)$ is a symmetric matrix (given in App.~\ref{app:MapToClockModel}, Eq.~\ref{eq:KernelsE2}) obtained from the phonon spectrum of the  WC with isotropic effective mass.

We have evaluated $\tilde{J}(\kbs)$ numerically and found that its largest eigenvalue over all $\kbs$ arises at the $\Mbs,\Mbs',\Mbs''$ points. The eigenvector at $\Mbs= \frac{2\pi}{\sqrt{3} }[1,0]$ is $[1,0]^{\top}$. Therefore, 
were we to ignore the discreteness of the phase variables, we would conclude that the optimal pattern would be $\ee^{2\ii \q_i} = \ee^{\ii \Rbf_i \cdot \Mbf}$ (or the analogous state at $\Mbs'$ or $\Mbs''$). This is thus a lower bound on the energy of the discrete model. Taking the constraint on values of $\theta_j$ into account, we can achieve this optimal pattern with $\theta_{WC}=0$ so long as the number of valleys, $n_v$, is even. The order for $n_v=2$ is shown in Fig.~\ref{fig:nv_2}. For $n_v$ odd, no state saturates the lower bound, but one can construct states that are close to the ideal state.  The case of $n_v=1$ (which is equivalent to the fully polarized state for arbitrary $n_v$) is treated in Sec.~\ref{sec:nv_1}.  For $n_v=3$ we can construct a close to optimal state with $\ee^{2\ii \q_i}= \frac{-1+\sqrt{3}\ii \ee^{\ii \Rbf_i\cdot \Mbs}}{2} $. We have checked that this configuration has lower energy than any state with no more than three electrons per unit cell, although we have not proven that it has the absolute minimum energy.
}
\subsection{Arbitrary mass anisotropy: Variational Approach}\label{sec:Variational}
\def\BZp{\text{BZ}'}

We study the $T=0$ phases at arbitrary mass anisotropy $\lambda$ using variational trial states. Specifically, we assume periodic pseudo-spin patterns with various sizes of unit cell. 

\begin{figure}
  \centering
      \includegraphics[width=\columnwidth]{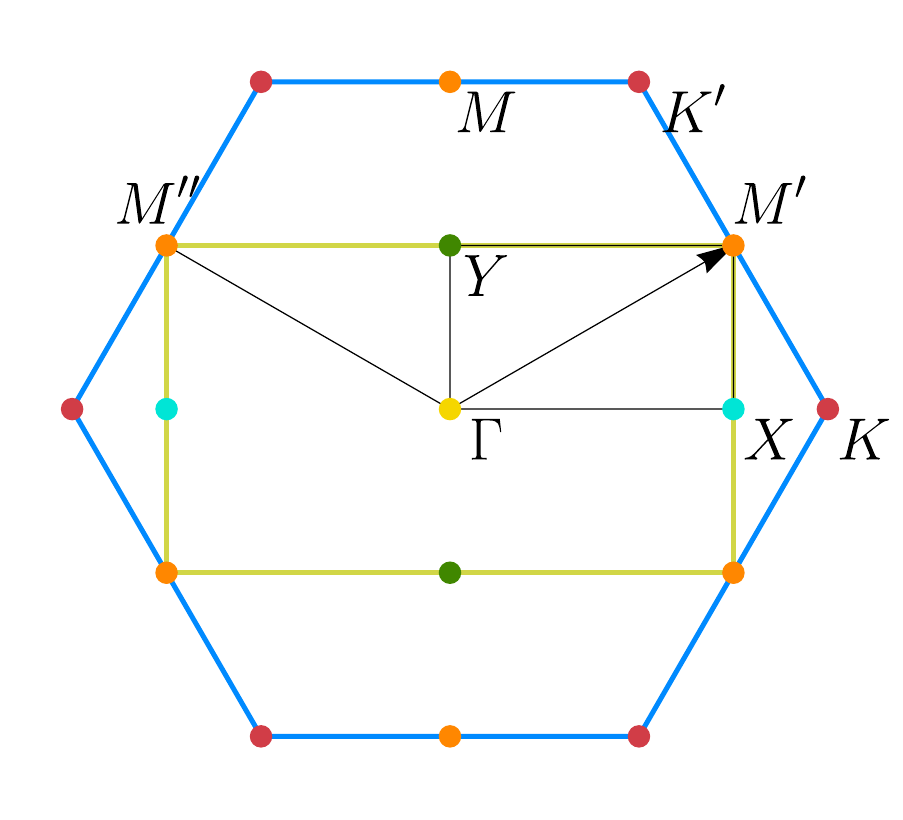}
    \caption{\label{fig:BZLabels} Wigner crystal's Brillouin zone (boundary in blue) and folded Brillouin zone  (boundary in yellow) for our candidate anisotropic Wigner crystals with period two stripe. (The black line corresponds to the path used in Fig.~\ref{fig:BandStructuresPhonon}.)}
\end{figure}

The energy per electron of each trial state can be written as
\begin{equation}\label{eq:EphononZeroPt}
    \begin{split}
        e_{\text{ph}}
        &=\frac{\nu}{2}\int_{\kbs \in \BZp}\frac{\dd \kbs^2}{(2\pi)^2}\Tr[\sqrt{\W_{\kbs}^2}]  
    \end{split}
\end{equation}
where $\BZp$ is the folded Brillouin zone with with size $1/B$, where $B$ is the number of sites on the new unit cell. The dynamical matrix is $\W^2_{\qbs} = \sqrt{\Wmc}\Kmc(\qbs)\sqrt{\Wmc}$. Here $\Wmc$ and $\Kmc(\qbs)$  are $2B\times 2B$ matrices that represent the inverse mass anisotropy and the elastic matrix within the pseudo-spin pattern unit cell. See App.~\ref{app:VariationalApproach} for details. For instance, the period-two stripe state from the small $\lambda$ limit has $B=2$, and $\BZp$ corresponds to the yellow rectangle in Fig.~\ref{fig:BZLabels}.

{For the three cases considered in Sec.~\ref{sec:appSemiconductors2d} ($n_v=1,2,3)$, we found that the following is a good fit for the energy of the preferred states (depicted in Figs.~\ref{fig:nv_1},\ref{fig:nv_2},\ref{fig:nv_3}, respectively):
\begin{equation}\label{eq:FitEph}
        \eph^{(n_v)}(\eta)_{\text{fit}} = \sqrt{\frac{2}{1+\h^2}} \left(A_0+A_1\h + A_2 \h^2+A_3 \h^3\right)
\end{equation}
where $\eta = \sqrt{\frac{m_L}{m_T}} \in [0,1)$. The coefficients $A_j$ are $n_v$ dependent. Their values are displayed in Tab.~\ref{tab:AparamsFit}}.

\begin{figure}[h]
\includegraphics[width=0.8\columnwidth]{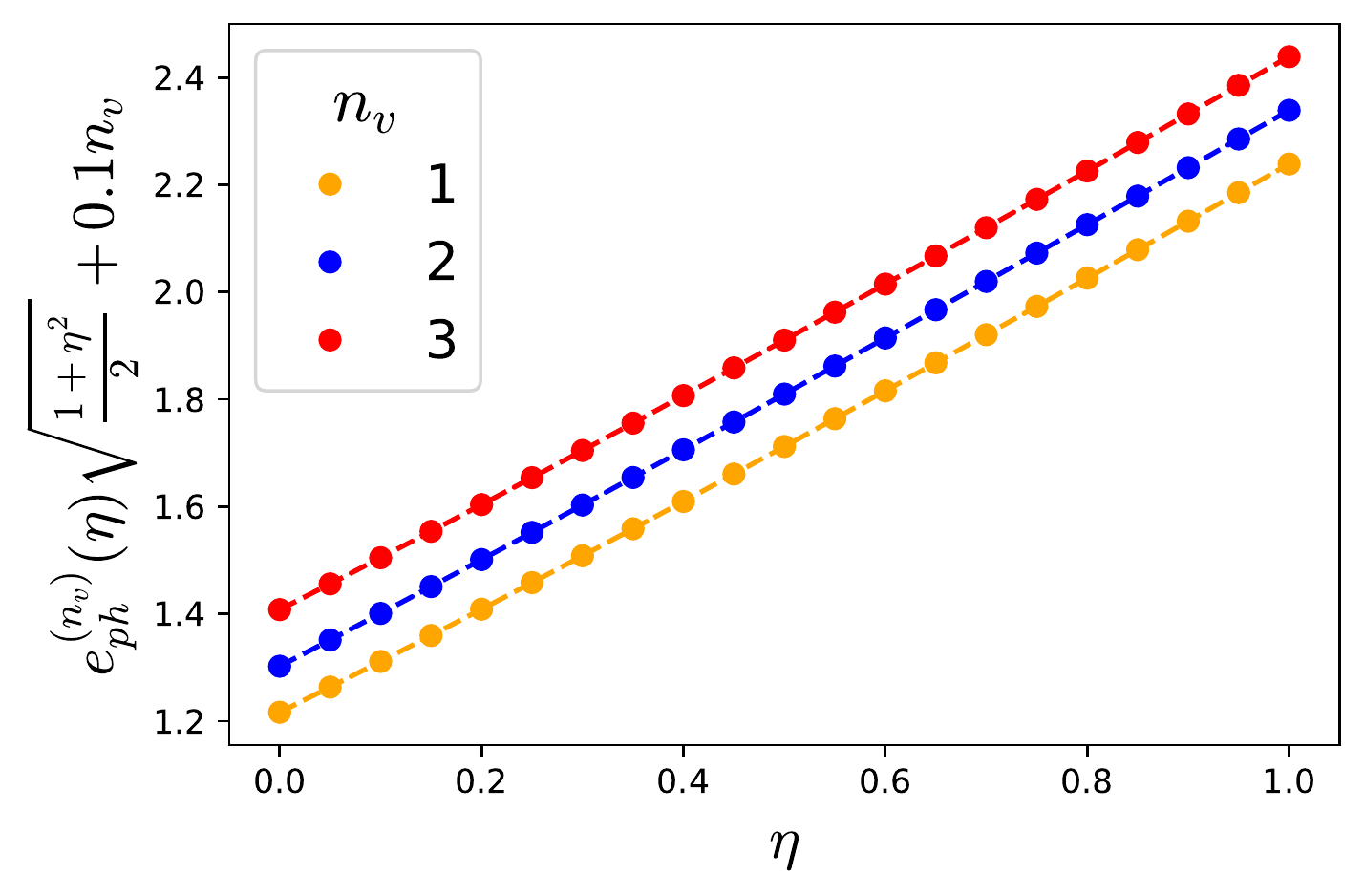}
\caption{\label{fig:energy_fit_together} Dependence of the normalized zero point phonon energy (Eq.~\ref{eq:EphononZeroPt}) for the different candidate states with different number of valleys ($n_v$). The dots are obtained via numerical evaluation of the integral and the dashed line corresponds to the fit in Eq.~\ref{eq:FitEph}. A sketch of the states are shown in Figs.~\ref{fig:nv_1}($n_v=1$), ~\ref{fig:nv_2}($n_v=2$) and ~\ref{fig:nv_3}($n_v=3$). (Note that we have introduced a shift to make the three curves distinguishable.)}
\end{figure}





\subsection{Physical quantities from variational states}

We calculate the following measurable properties using the trial states:
\begin{itemize}
    \item Broken symmetries;
    \item The phonon's dispersion relations\footnote{See App.~\ref{app:DispersionRelations} for details on how we calculate the phonon dispersion relations near $\Gamma$.};
    \item The optical conductivity.
\end{itemize}

{In particular, an expression for the conductivity can be obtained  from the Kubo formula, 
\begin{equation}
    \s_{ab}(\w) = 
    \frac{4\pi}{A}\sum_{n} 
    \frac{\d(\w -\w_{n})}{\w}\mel{0}{J_a}{n} {\mel{n}{J_b}{0} }.
\end{equation}
where $A$ is the area of the sample and the sum is over excited states, $\ket{n}$, with energies $\omega_{n}>0$. {We use $\vec{J}=\sum_i \vec{J}_i$ for the current operator, where $\vec{J}_i = e \vec{v}_{i}= e W_i\cdot \vec{p}_i$ is the contribution to the current of the electron at site $i$. After some algebra (see App.~\ref{app:OpticalConductivity}), the optical conductivity can be written as the sum of the contribution of each optical phonon with crystal momentum zero (in the reduced BZ). Note that the above calculation is equivalent to the computation of the optical conductivity using the Kubo formula according to  $\sigma(\omega) \equiv \lim_{\vec q \to \vec 0}\sigma(\vec q,\omega)$). 
This explains why we do not get a contribution from the acoustic modes, {i.e.}, there is no $\d(\omega)$ contribution.  (In the presence of disorder, this acoustic contribution would appear as a low frequency feature associated with a ``pinning mode.'') }

}

\subsection{Applications to 2d semiconductors}\label{sec:appSemiconductors2d}
\begin{table}
    \begin{tabular}{|c||c|c|c|c||c|}
    \hline
        $n_v$ & $A_0$& $A_1$& $A_2$ & $A_3$ & Fig. $\#$ \\\hline
        $1$ & 1.116 & 0.941 & 0.120 & -0.0395 & \ref{fig:nv_1}\\
        $2$ & 1.102 & 0.979 & 0.083 & -0.0267 & \ref{fig:nv_2}\\
        $3$ & 1.108 & 0.962 & 0.105 & -0.0360 & \ref{fig:nv_3} \\ \hline
    \end{tabular}
    \caption{\label{tab:AparamsFit} Fit parameters (Eq.~\ref{eq:FitEph}) for the variational energy $\eph^{(n_v)}(\eta)_{\text{fit}} = \sqrt{\frac{2}{1+\h^2}} \left(A_0+A_1\h + A_2 \h^2+A_3 \h^3\right)$ ( $\eta =\sqrt{ \frac{m_L}{m_T}} \in [0,1)$) of the states depicted in the respective figure.}
\end{table}
\subsubsection{One valley \texorpdfstring{$(n_v=1)$}{nv=1}}\label{sec:nv_1}

\begin{figure}[h]           
\includegraphics[width=0.8\columnwidth]{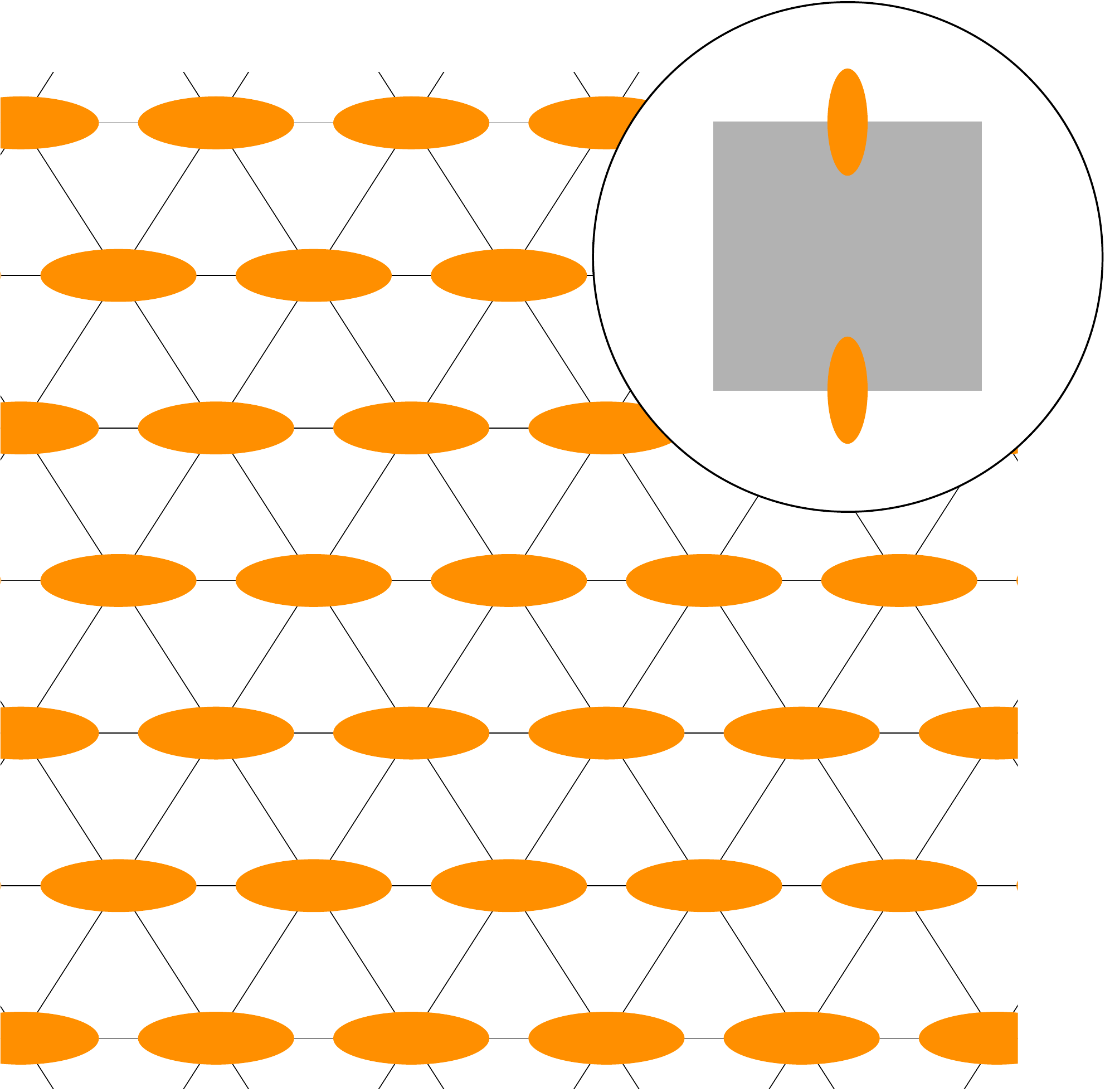}
\caption{\label{fig:nv_1} Schematic real space charge density in a WC with a single anisotropic valley ($n_v=1$). {(Inset shows the orientation of the underlying crystal's BZ with respect to the WC.) } }
\end{figure}

We start by considering the case of a single valley  with an anisotropic effective mass, which is also relevant to the case of a system with multiple valleys in a fully valley polarized state. 
%
In this case, 
we only need to consider a unit cell with a single site. We find that the optimal orientation of the WC relative to the principal axes of the effective mass tensor is such that the long direction of the Fermi contour is perpendicular to one of the WC's axes. A depiction of the 
charge density distribution that results from the electron zero-point motion is shown in Fig.~\ref{fig:nv_1}. A fit and plot for the energy of this variational state can be found in Tab.~\ref{tab:AparamsFit} and Fig.~\ref{fig:energy_fit_together}, respectively. 




The phonon dispersion near $k=0$ 
is \textit{anisotropic}:
\begin{equation}
    \begin{split}
        \w_{ac,L}(\kbs) 
        &= \sqrt{k\kWC\lrRb{\cos[2](\q_{\kbs}) +\eta^2\sin[2](\q_{\kbs})}}\\
        \w_{ac,T}(\kbs) &=  \frac{\mathrm{v} k}{\sqrt{\cos[2](\q_{\kbs}) +\eta^2\sin[2](\q_{\kbs})}}.
    \end{split}
\end{equation}
where $\mathrm{v}^2 \approx 0.245 (1-\l^2)$ and $\kbs=[k\cos(\q_{\kbs}),k\sin(\q_{\kbs})]$.
{ Along symmetry directions, the lower branch is the transverse acoustic mode and the upper is the longitudinal acoustic mode, so we more generally label them $L$ and $T$.}
The dispersion along various symmetry directions is displayed in Fig.~\ref{fig:dispersion_nv_1}. There is no optic phonon mode, and consequently no optical response. 

\subsubsection{Two valleys related by \texorpdfstring{$C_4 (n_v=2)$ }{C4}}\label{sec:nv_2}

\begin{figure}[h]           
\includegraphics[width=0.8\columnwidth]{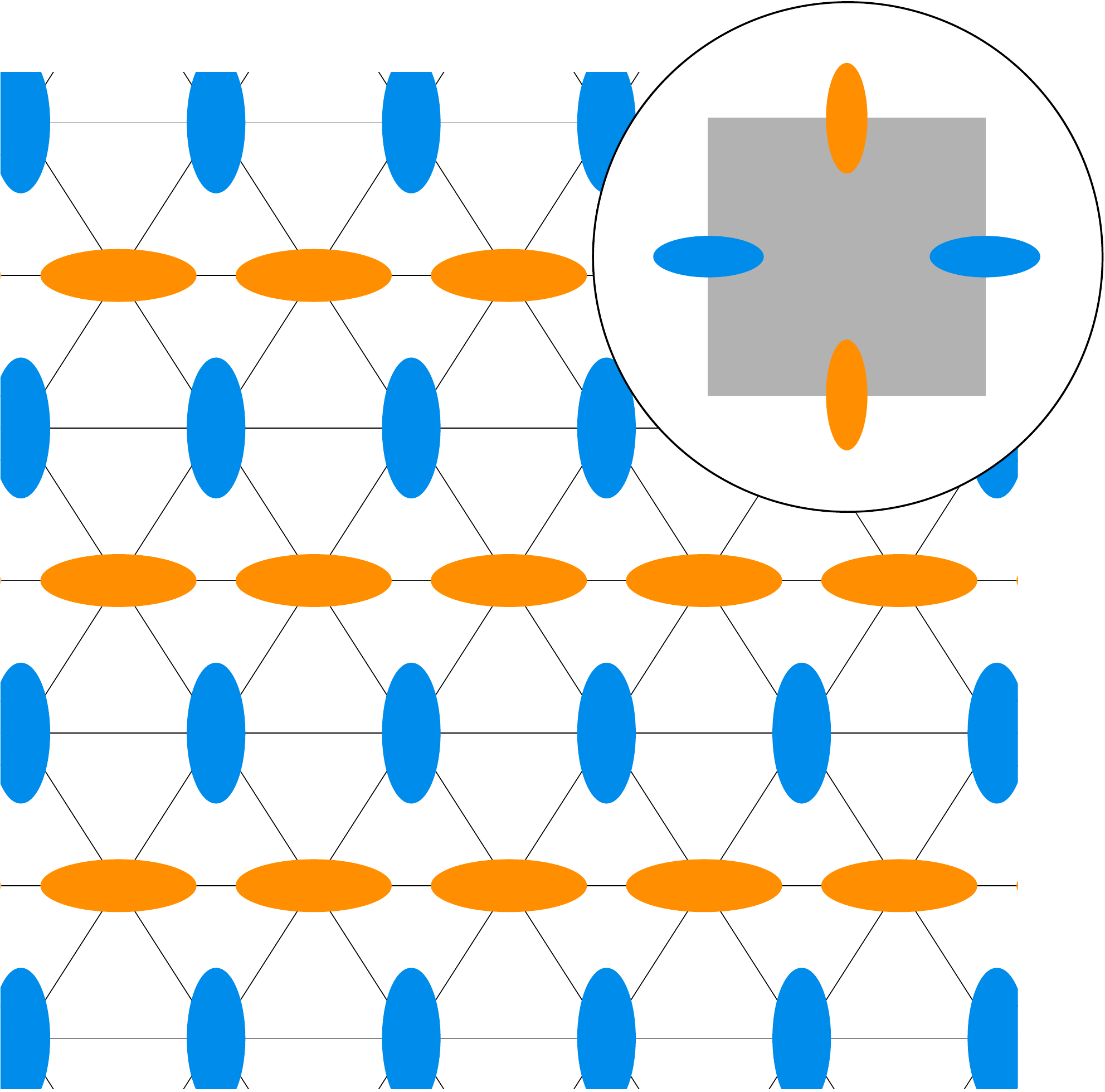}
\caption{\label{fig:nv_2} Schematic real space charge density in a WC with two valleys related by $C_4$ symmetry ($n_v=2$). {(Inset shows the orientation of the underlying crystal's BZ with respect to the WC.) } }
\end{figure}


In this section we consider a semiconductor with $PG=D_4$ and two pockets at 
$X=[\pi,0]$ and $Y=[0,\pi]$, {as shown in Fig.~\ref{fig:valleys}-{\color{blue} a}}. The two valleys are related by $C_4$ symmetry. We consider $\eta^2=m_T/m_L\in [0,1)$. {This section could be relevant to AlAs \cite{Hossain2021SpontaneousValleyPolarization}. }

We have computed $\eph$ for all symmetry inequivalent valley ordered patterns with unit cells of up to six sites. The valley pattern with the lowest $\eph$ has the period-two valley-stripe order depicted in Fig.~\ref{fig:nv_2}. A close competitor is the fully polarized state considered above. In particular, the ratio $\frac{e^{(2)}_{\text{ph}}-e^{(1)}_{\text{ph}}}{e^{(2)}_{\text{ph}}+e^{(1)}_{\text{ph}}}$ depends on the mass anisotropy and varies from $0$ to approximately $6.4\times 10^{-2}$. Our results coincide with the results of $\eph$ from Ref.~\cite{PhysRevB.89.155103} in the limit $\eta\rightarrow 0$, and with our results in Sec.~\ref{sec:SmallMassAniPerturbative} in the $\eta \to 1$ limit. 

There is an equal number of $X$ and $Y$ valley electrons per unit cell so there is no net valley polarization. However, the PG is broken to $D_2$ because the two flavours are at inequivalent positions in the unit cell.

\begin{figure}[h]
\subfloat[\label{fig:dispersion_nv_1}]{
  \centering
    \includegraphics[width=.8\columnwidth]{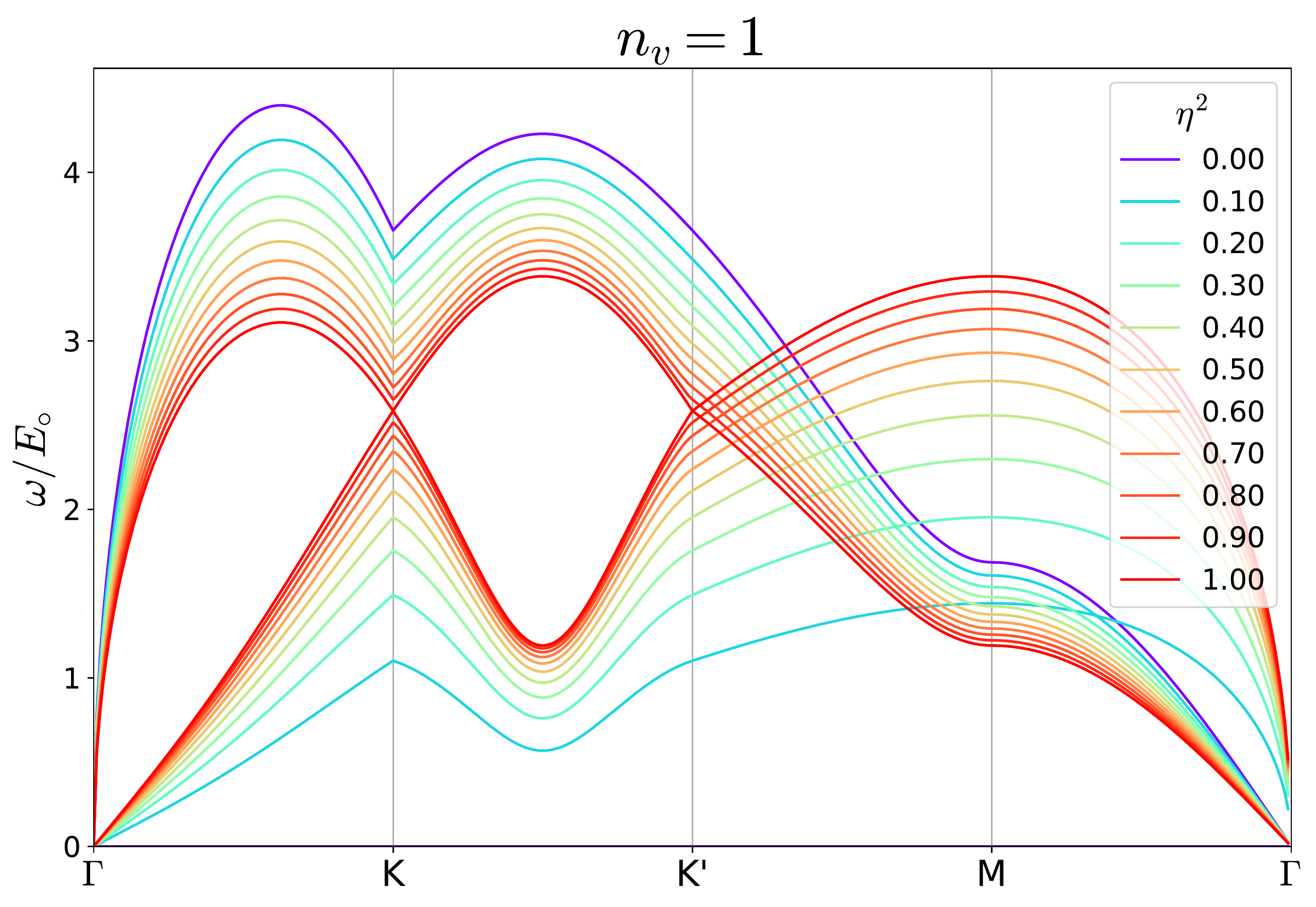}
}\\
\subfloat[\label{fig:dispersion_nv_2}]{
  \centering
    \includegraphics[width=.8\columnwidth]{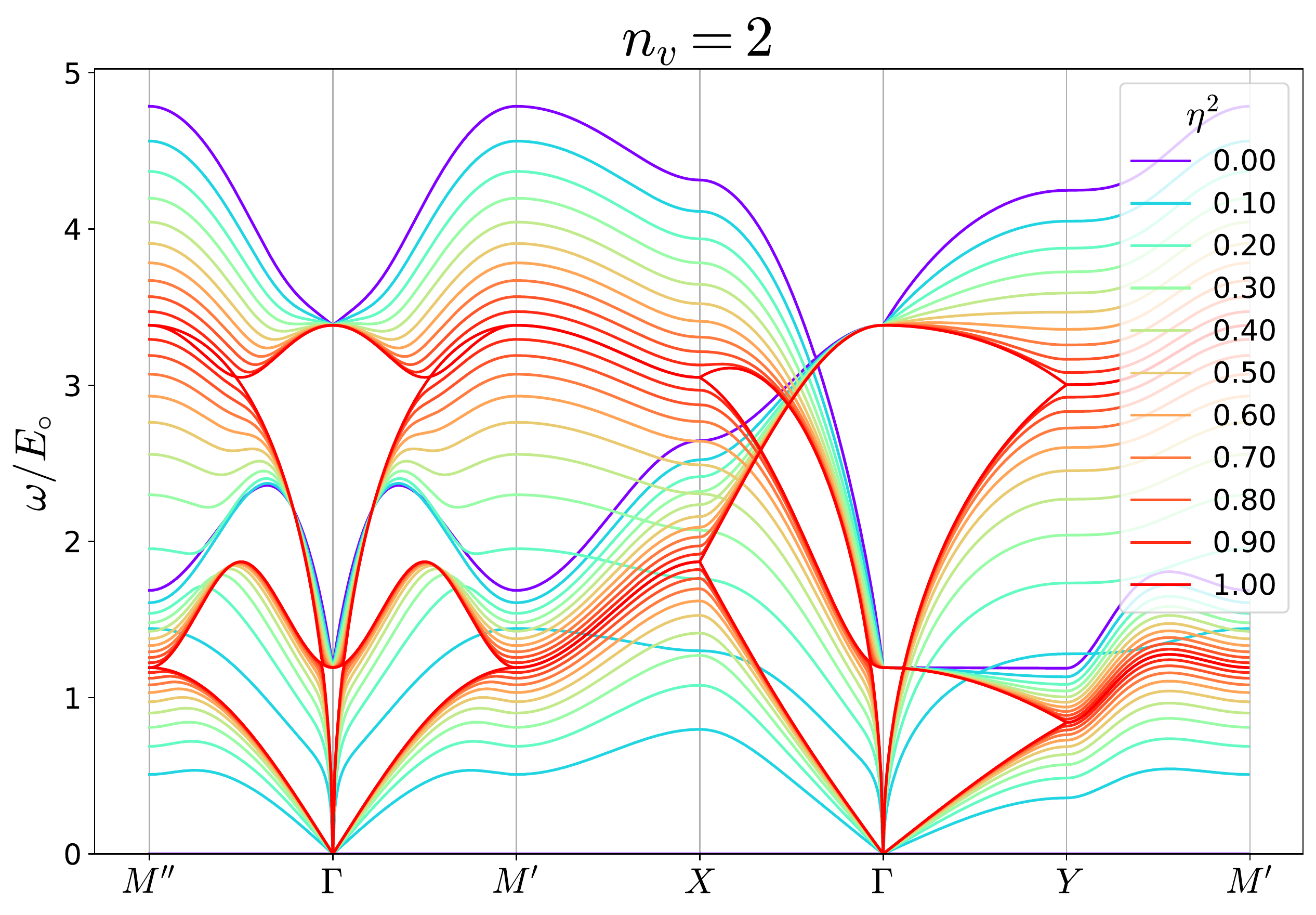}
}\\
\subfloat[\label{fig:dispersion_nv_3}]{
  \centering
    \includegraphics[width=.8\columnwidth]{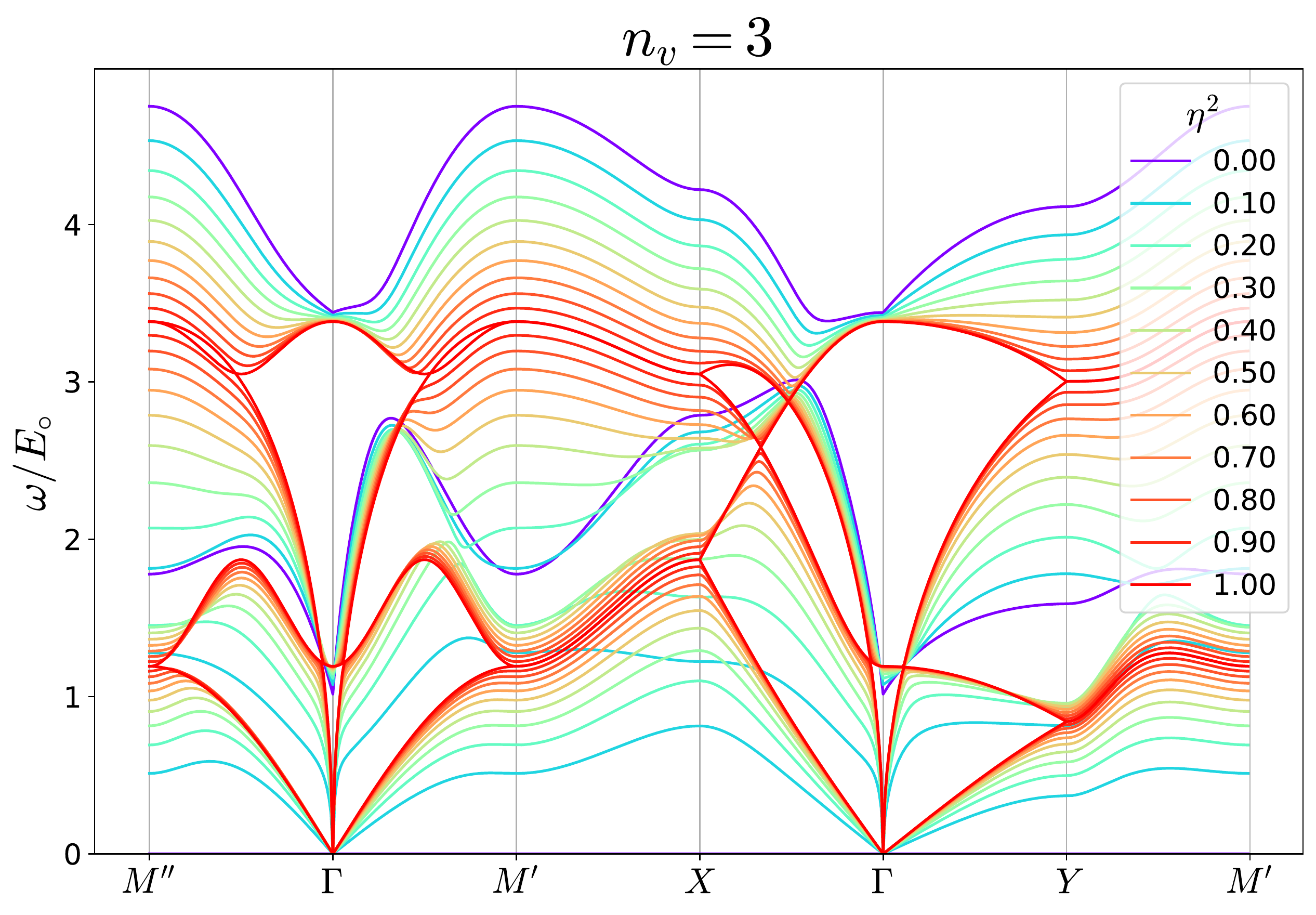}
}
\caption{\label{fig:BandStructuresPhonon}Phonon band structure for the three relevant states for various values of $\eta^2= \frac{m_T}{m_L}$. See Fig.~\ref{fig:BZLabels} for the meaning of the momenta labels. The polarized state ($n_v=1$) has one electron per unit cell, and so two acoustic branches, while the stripe states ($n_v=2,3$) have two electrons per unit cell, and hence two additional optic modes. 
}
\end{figure}

As there are two electrons per unit cell, there are 4 phonon branches - two optical and two acoustic. Even though the valley pattern has only $D_2$ symmetry, the long wave-length acoustic modes are \textit{isotropic}
\begin{equation}
    \begin{split}
        \w_{ac,L}(\kbs) 
        &=  \sqrt{\left(1-\l^2\right)k\kWC }\\
        \w_{ac,T}(\kbs) 
        &= \mathrm{v} k \\
    \end{split}
\end{equation} where $\mathrm{v}^2\approx 0.245(1-\l^2)$ and $\kWC= \frac{2\pi}{ \sqrt{\nu}}$.


{In Fig.~\ref{fig:dispersion_nv_2} we show the phonon dispersion relations along  high-symmetry directions for several values of $\eta^2=m_T/m_L$. Note that the phonon frequencies at $\kbs=\Gamma$ are independent of $\eta$.}


The optical conductivity  for the state depicted in
Fig.~\ref{fig:nv_2}
is the sum of contributions from the two optic modes,  $\s(\w)= \s^{(1)}(\w) + \s^{(2)}(\w)$:
\begin{equation}
    \begin{split}
    \s_{ab}^{(1)}(\w)  &= \s_0 \l^2 \left[\d_{ax}\d_{bx}\d(\w-\w_1)\right] \\
    \s_{ab}^{(2)}(\w)  &=  \s_0 \l^2 \left[\d_{ay}\d_{by}\d(\w-\w_2)\right]\\
    \w_{1} &\approx 1.19255 \\
    \w_{2} &\approx 3.38394, 
\end{split}
\end{equation}
$\s_0 =\frac{2\pi e^2 \nel}{\mst} \frac{1}{\Enot}$ and $\w$ is measured in units of $\Enot$,
and the imaginary part can be computed straightforwardly from Kramers-Kronig.  Note that the existence of optic modes is directly related to the doubling of the WC unit cell; they would not arise, for example, in a fully valley polarized state.



\subsubsection{Three valleys related by \texorpdfstring{$C_3$}{C3}}\label{sec:nv_3}

\begin{figure}[h]
    \includegraphics[width=0.8\columnwidth]{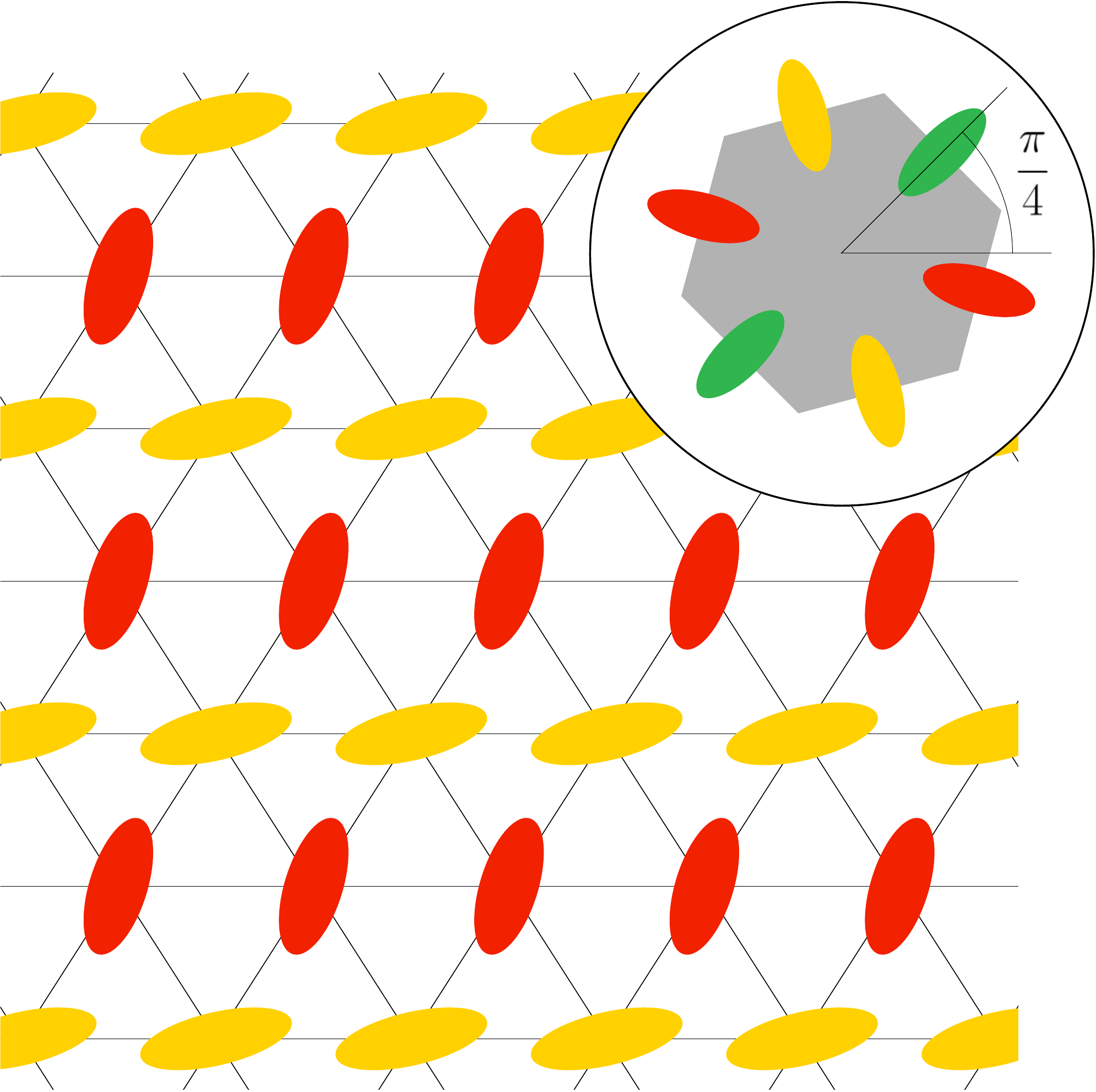}
    \caption{\label{fig:nv_3} Schematic charge density in a period-2 stripe valley ordered WC with three valleys $(n_v=3)$ related by $C_3$ symmetry. {(Inset shows the orientation of the underlying crystal's BZ with respect to the WC.) }  }
\end{figure}

Next, we consider a system with $PG=D_6$ with three valleys related by threefold rotations. For instance, the Fermi pockets could be at the $M$ points (the middle points of the edge of the hexagonal BZ), {as shown in Fig.~\ref{fig:valleys}-{\color{blue} b}}.

We considered unit cells with up to 3 electrons per unit cell. We find that the configuration with the lowest energy is a period-2 valley stripe pseudo-spin antiferromagnet. 
The fully polarized state is again a close competitor.

The valley order has momentum $M$ (as for $n_v=2$) but only electrons in two valleys out of the three valleys participate.  The axes of the Wigner crystal no longer align with those of the underlying lattice. Instead, the horizontal axes of the WC lie preferentially at about $15\degree$ relative to one principal axes of the underlying crystal.  Equivalently, the direction of the principal axes of the effective mass tensor of the unpopulated valley is oriented at $45\degree$ with respect to the horizontal axes of the WC. A cartoon 
of this
state is shown in Fig.~\ref{fig:nv_3}. The orientation is the equivalent to the proposed low energy state for the effective XY model $n_v=3$ of Sec.~\ref{sec:SmallMassAniPerturbative}.

The long wave-length longitudinal (L) and transverse (T) acoustic modes
are \textit{anisotropic},
\begin{equation}
    \begin{split}
        \w_{ac,L}(\kbs) 
        &= \mathrm{v}_L \sqrt{k\kWC}\sqrt{1+\frac{\l}{2} \sin(2\varphi) }  \\
        \w_{ac,T}(\kbs) 
        &= \frac{\mathrm{v}_R k}{\sqrt{1+\frac{\l}{2} \sin(2\varphi)}}
    \end{split}
\end{equation}
where $\mathrm{v}_{L/R}\propto \sqrt{1-\l^2}$ and $\kbs= k  [\cos(\varphi),\sin(\varphi)]^{\top}$.

{In Fig.~\ref{fig:dispersion_nv_3} we show the phonon dispersion relation along  high-symmetry directions for several values of $\eta^2=m_T/m_L$. Note here that the optical phonon frequencies at $\kbs=\Gamma$ are $\eta$ \textit{dependent}.}

\begin{figure}[h!]
\subfloat[\label{fig:opticalFrequencies_M=3}]{
  \centering
    \includegraphics[width=0.8\columnwidth]{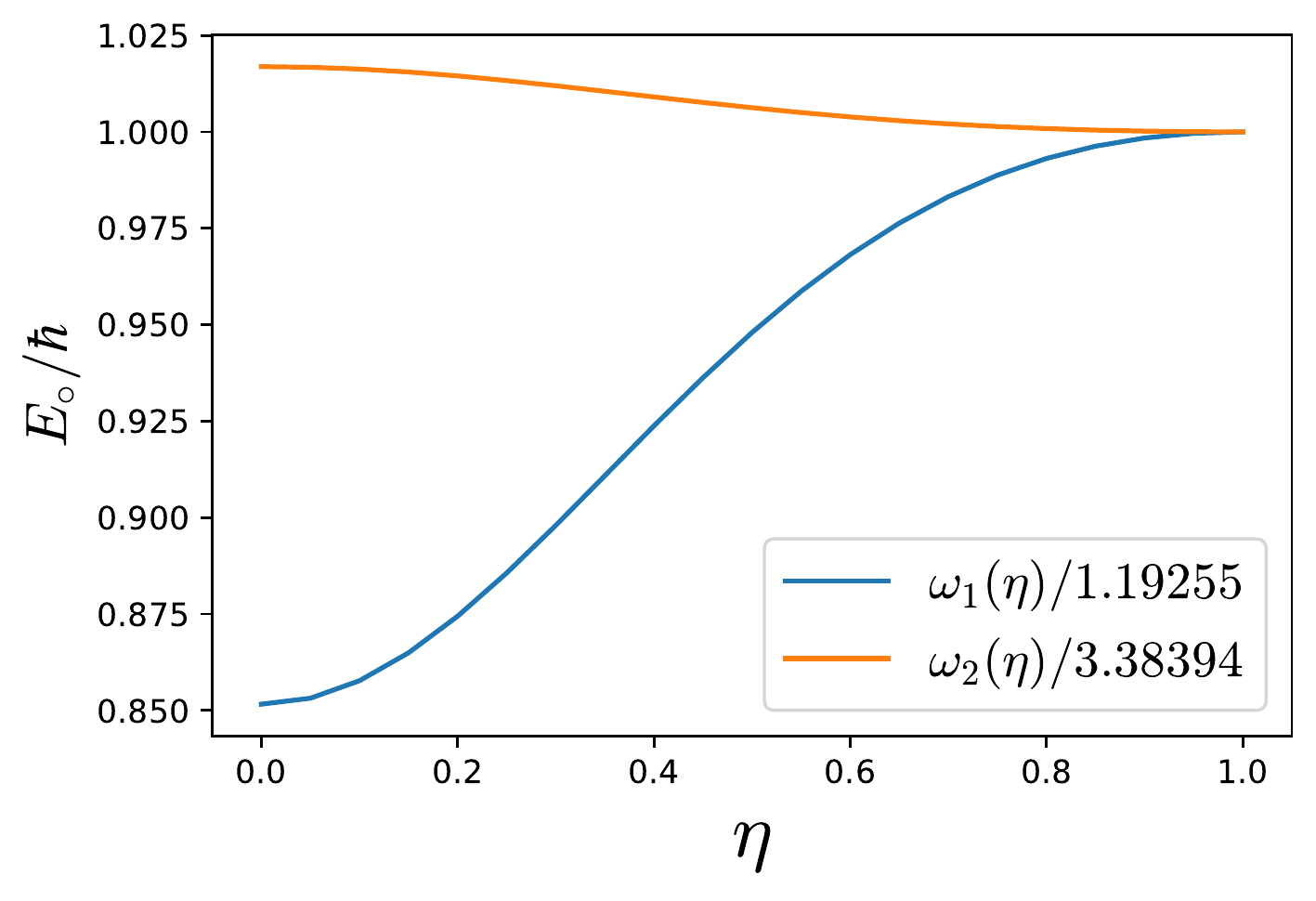}
}\\
\subfloat[\label{fig:transmissionAngles_M=3}]{
  \centering
\includegraphics[width=0.8\columnwidth]{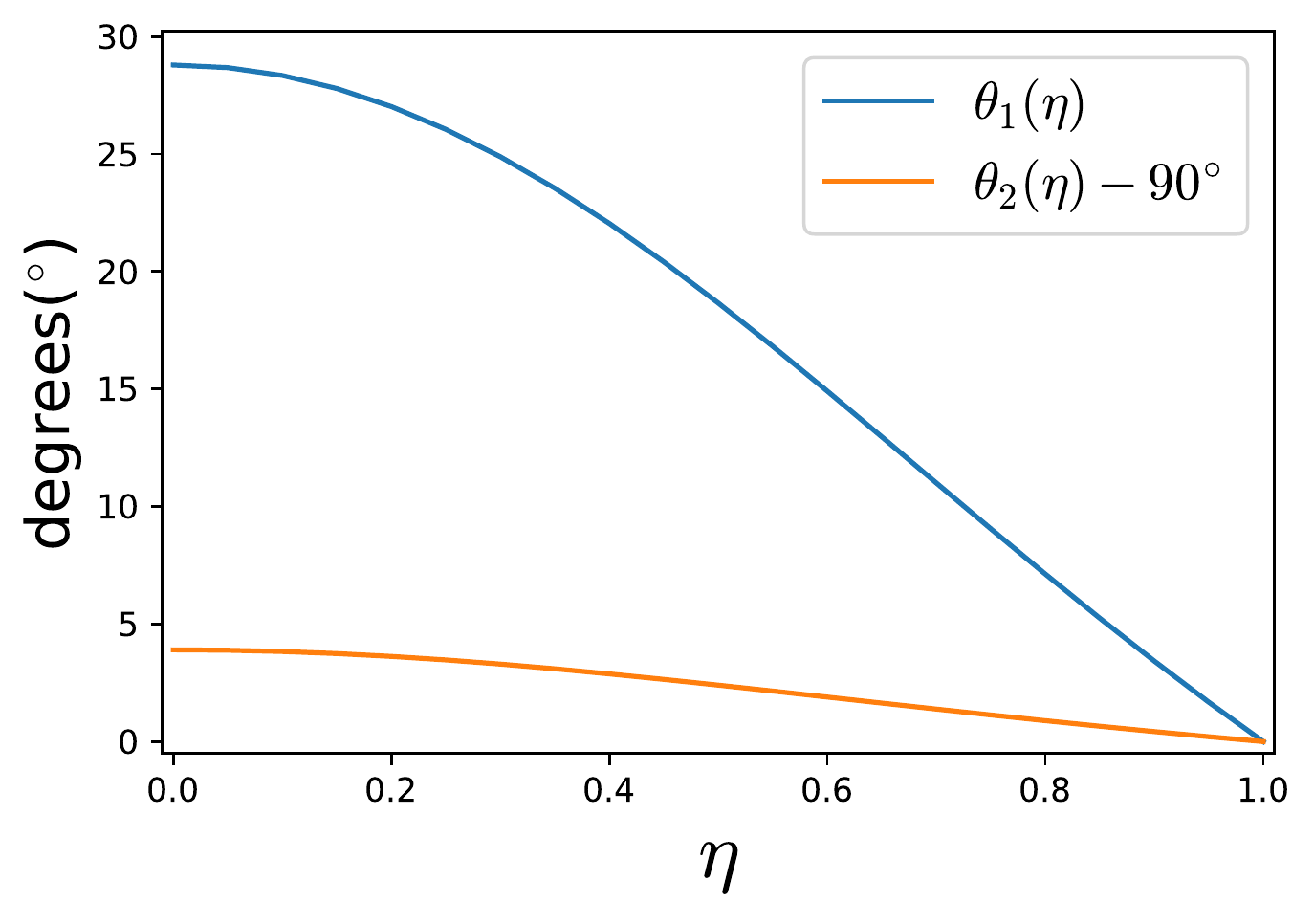}
}\\
\subfloat[\label{fig:matrixElements_M=3}]{
  \centering
\includegraphics[width=0.75\columnwidth]{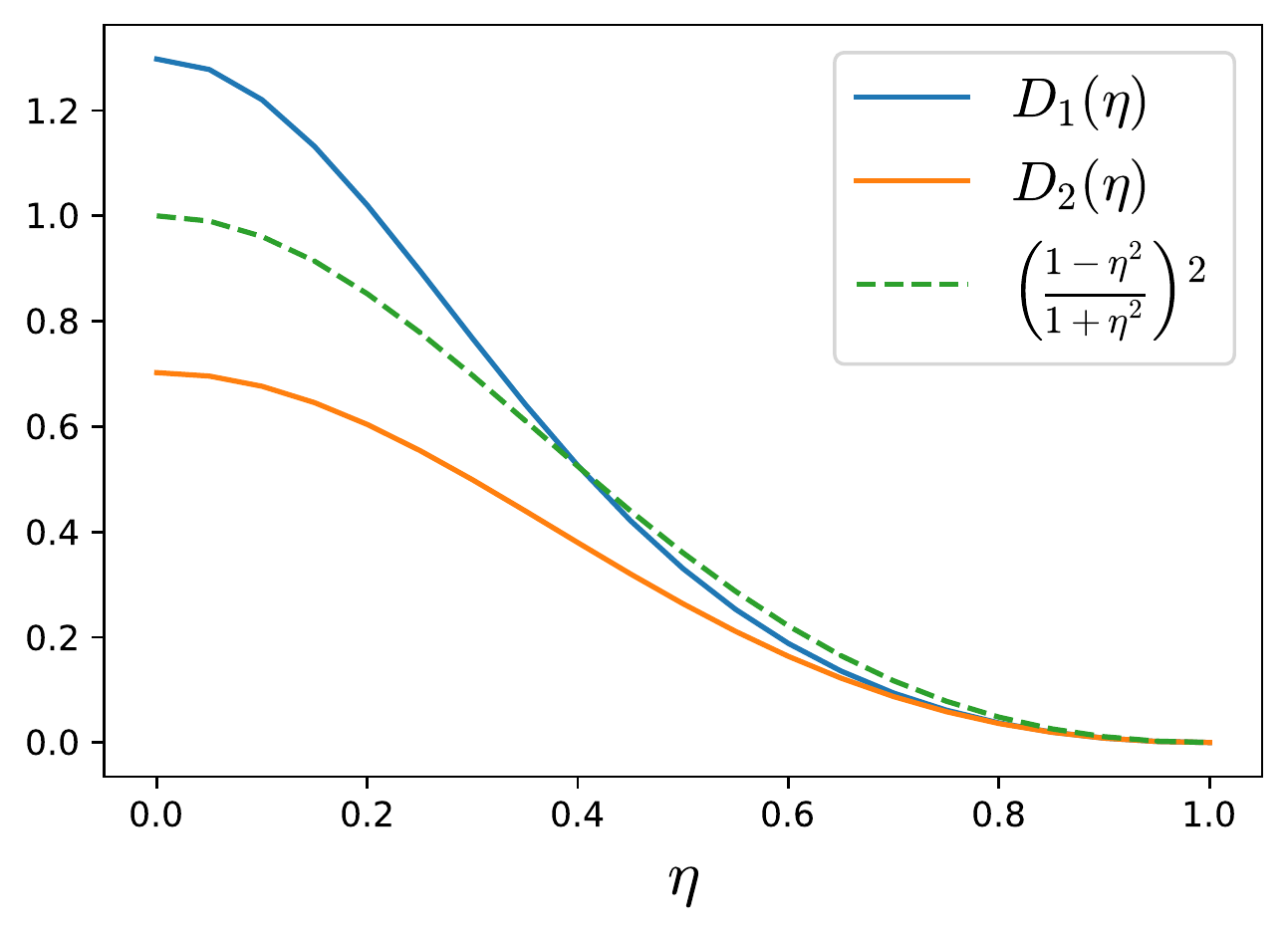}
}
\caption{\label{fig:etaDependenceOptical_M=3} Dependence of parameters of the optical conductivity of stripe phase for $n_v=3$. ${ (a)}$ phonon frequency normalized with respect to the $\eta=1$ model ($E_{\circ}= \sqrt{\frac{e^2}{4\pi \eps a^3 \mst }}$). ${(b)}$ Deviation of the transmission axes from the isotropic limit $(\eta\to 1^-)$. ${(c)}$ weight of optical conductivity for the optical bands. The green line is a guide for comparison to the weight of optical conductivity of the $n_v=2$ state.  }
\end{figure}

This pseudo-spin order has inversion symmetry but lacks reflection symmetry. This shows up in the optical response as an $\eta$ dependence of the transmission axes. The optical frequency is the sum of contributions of the two optical modes $\s(\w)=\s^{(1)}(\w) + \s^{(2)}(\w)$,
\begin{equation}
    \begin{split}
    \s_{ab}^{(j)}(\w)  &= \s_0 D_{j} \left[v^{(j)}_{a}v^{(j)}_{b}\d(\w-\w_{j})\right] 
\end{split}
\end{equation}
here $\s_0 =\frac{2\pi e^2 \nel}{\mst} \frac{1}{\Enot}$, $\omega_{j}$ is the frequency of optical mode $j$,  $v_{a}^{(j)}=[\cos(\q_j), \sin(\q_j)]^T$ is a polarization vector, and $D_{j}$ is the weight of the delta function. The $\eta$ dependence of $\w_{j},\q^{(j)}, D_{j}$ is plotted in Fig.~\ref{fig:etaDependenceOptical_M=3}. 


\section{Inversion symmetry breaking }\label{sec:InversionPairs}
In this section we consider possible effects that could break the degeneracy between pairs of valleys related by inversion symmetry. For this section, one should have in mind valleys at the $\pm \Kbs$ points of a crystal with $C_6$ symmetry, {as shown in Fig.~\ref{fig:valleys}-{\color{blue} c}}. Our considerations may be relevant for monolayer and bilayer transition metal dichalcogenides \cite{smolenski2021signatures, zhou2021bilayer}.

\subsection{{Trigonal warping}}\label{sec:cubicCorrection}

We start by considering trigonal warping, {i.e.} we go beyond the effective mass approximation and add a cubic in $\pbs$ term to the kinetic energy in Eq.~\ref{eq:KineticEnergyEffectiveMass}:
\begin{equation}
    T_{\alpha}(\pbs) = \sum_{a,b}\frac{1}{2}W_{ab}^{(\a)}p_ap_b + \sum_{a,b,c} \frac{1}{3!}t^{(\a)}_{abc}p_ap_bp_c
\end{equation}
If valleys $\a,\a'$ are related by inversion, we must have $W^{(\a)}=W^{(\a')}$ and $t^{(\a)}=-t^{(\a')}$.  { Needless to say, in order to insure that the Hamiltonian is bounded below, we need to include higher order terms as well - at least to order $p^4$.  However, since we will treat the effects of trigonal warping  in perturbation theory, we do not need to include these higher order terms explicitly.} 

For simplicity, we consider a pair of $C_3$-symmetric valleys related by inversion. The dispersion relation with $p_x=p \cos(\theta_{\pbs})$ and $p_y=p\sin(\theta_{\pbs})$ simplifies to 
\begin{equation}
    T_{\a}(\pbs) = \frac{p^2}{2m}+ t p^3 \cos(3\q_{\pbs}-3\phi_{\alpha}),
\end{equation}
where $\phi_{\alpha}$ is an angle that fixes the orientation of the Fermi contour with respect to Wigner crystal coordinates. If $\a$ and $\a'$ label a pair of valleys related by inversion, then $3\phi_{\a'}=3\phi_{\a}+\pi$. We rescale the position and momentum variables as in  Sec.~\ref{sec:EffectiveMassHarmonicApprox} and use the harmonic approximation. The effective Hamiltonian in rescaled variables is
\begin{equation}\label{eq:HTrigonalwarping}
     \begin{split}
         h_{\eff}= &\sum_{i} \frac{1}{2}P_{i}^2 +\sum_{i, j} \frac{1}{2} Q_iK_{ij}Q_j\\
         &+\eps_{3}\sum_{i}[\ee^{3\ii \phi_{\a_i}}(P_{i;x} - \ii P_{i;y})^3 + h.c.]
     \end{split}
\end{equation}
where $\eps_3=t\sqrt{\Enot (\mst)^{3}}$ goes like $\eps_3 \propto \rs^{-3/4}$ as $\rs\to\infty$. Due to the $C_3$ symmetry, the next leading angle-dependent correction to the effective mass approximation is at order $P^6$ which has a prefactor that scales as $1/\rs^{3}$.

The details of the calculations can be found in App.~\ref{app:C2breaking}. Here we give an overview of the results. As we did in Sec.~\ref{sec:SmallMassAniPerturbative}, we expand the phonon energy with trigonal warping $e^{\triangle}_{\text{ph}}=\sum_{n=0}^{\infty}e_n^{\triangle} \eps_3^n$ where each $e_n^{\triangle}$ in principle can depend on the pattern of valley ordering. We find that $e_0^{\triangle}$ is independent of the ordering pattern. $e_0^{\triangle}+\eps_3e_1^{\triangle}$ is an effective potential for $P_{CM}=\frac{1}{\Nel}\sum_{i}P_i$. Upon including a quartic term in $P_{CM}$ for stability, the potential is optimized by $P_{CM}=0$. (Were we to consider a system whose $\eps_3 =0 $ ground state lacks $C_3$ rotation symmetry, there could be a term $\eps_3 P_{CM}$ in the effective potential so there would be a non-zero $P_{CM} \propto \eps_3 $. We can interpret this as a displacement of the band minima in the BZ of the underlying crystal.)

\begin{figure}[h]
    \includegraphics[width=0.9\columnwidth]{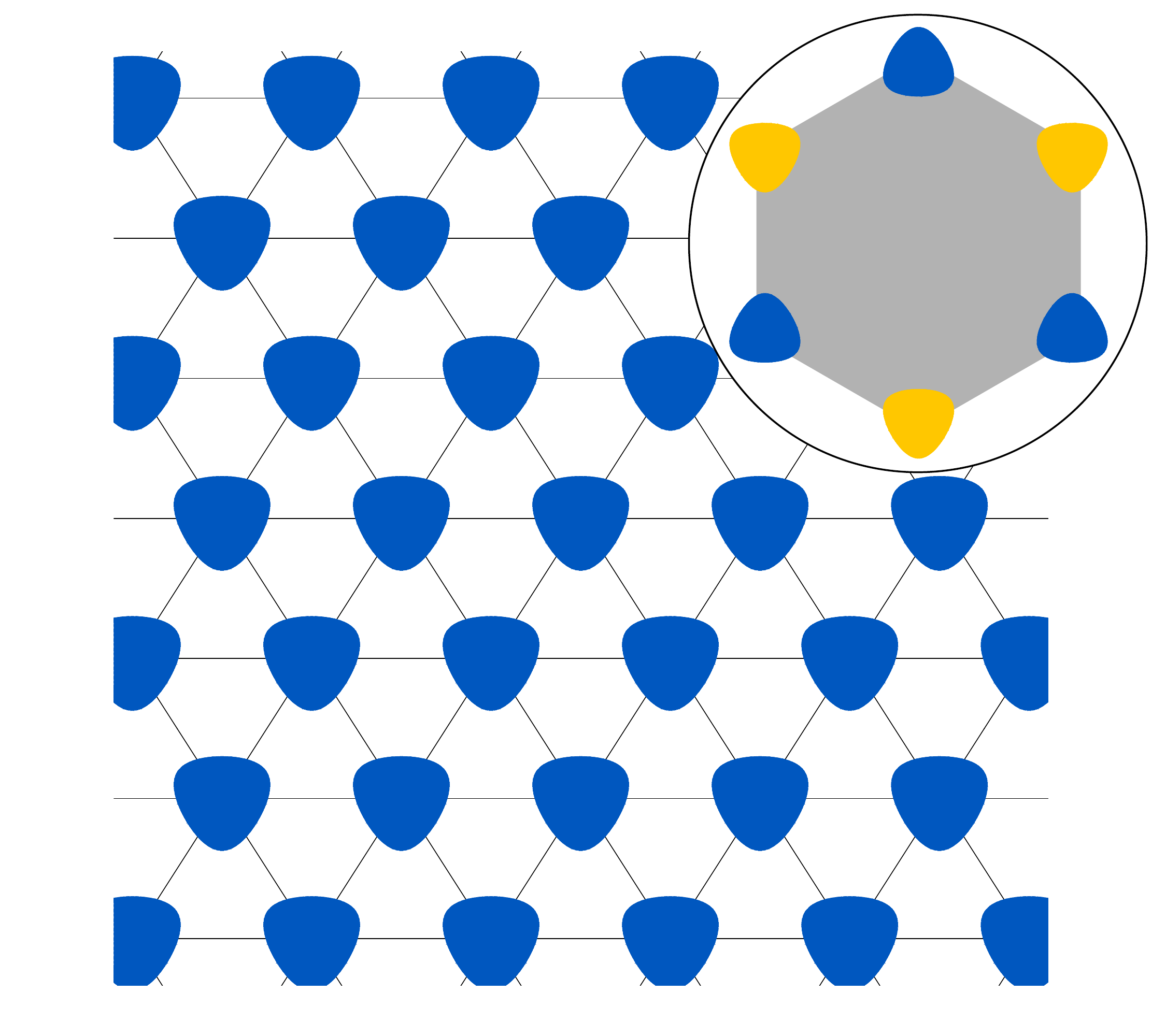}
    \caption{\label{fig:trigonalReal}
    Schematic real-space charge density of the valley-polarized state for a 2DEG with two valleys related by inversion symmetry at the $\pm \Kbs$ points and with trigonal warping. (Inset shows the orientation of the underlying crystal's BZ with respect to the WC)  }
\end{figure}

The next order correction to the energy is $\eps_3^2 e_2^{\triangle}$ with 
\begin{equation}\label{eq:SecondOrderCorrection}
   e_2^{\triangle}
   =-\frac{1}{\Nel}\sum_{\kbs} \Phi^{\dag}_{3;\kbs}\tilde{J}_3(\kbs)\Phi_{3;\kbs}^{},
\end{equation}
where ${\Phi}_{3,\kbs}$ is the Fourier transform of $[\cos(3\phi_i),\sin(3\phi_i)]^{\top}$ and $\tilde{J}_{3}(\kbs)$ is a symmetric matrix (given in App.~\ref{app:C2breaking}, Eq.~\ref{eq:KernelTRS}) obtained from the phonon spectrum of the  WC with isotropic effective mass. The largest eigenvalue of $\tilde{J}_{3}(\kbs)$ is at $\kbs=0$ with eigenvector $[0,1]^{\top}$. Therefore, the state with lowest energy is a valley-ferromagnet ($\kbs=0$) and $3\phi_{i}$ is constant and equal to $\pm \frac{\pi}{2}$, {as shown schematically in Fig.~\ref{fig:trigonalReal}}. 

\subsection{Berry curvature}

The Berry curvature can be thought of as a magnetic field in momentum space. Therefore, in a semi-classical picture (like ours) we need to modify the commutation relations between the physical position and momentum variables ( see e.g. Refs.~\cite{Chang2008BerryCurvatureEMFieldsRev,Qian2010BerryPhaseReview}) to 
\begin{equation}\label{eq:EqBerryComm}
\begin{split}
    \{r_a^{\text{phys}},r_b^{\text{phys}}\} &= \e_{ab}\Bmc; \\
    \{p_a^{\text{phys}},p_b^{\text{phys}}\} &= 0;\\
    \{r_a^{\text{phys}},p_b^{\text{phys}}\} &= \delta_{ab}. \\
\end{split}    
\end{equation}
where {
$\e_{ab}$ is the Levi-Civita symbol.} Here $\Bmc$ is the Berry curvature, assumed to be constant for simplicity. We identify the canonical variables
\begin{equation}
    \begin{split}
        r_{a}^{\text{canon}} &= r_{a}^{\text{phys}} + \frac{1}{2}\Bmc \e_{ab}p^{phys}_b,\\
    p_a^{\text{canon}} &= p^{phys}_a.
    \end{split}
\end{equation}
that are used to quantize the theory.

At large $\rs$, the Coulomb energy dominates so the electrons form a triangular lattice. Next, we use the harmonic approximation as in Sec.~\ref{sec:EffectiveMassHarmonicApprox} to get the same Hamiltonian as Eq.~\ref{eq:HamiltonianPhonon} with $Q_i$ and $P_i$ replaced by $Q_i^{\text{phys}}= {\sqrt{\Enot\mst}}q^{\text{phys}}_{i}$ and $P_i^{\text{phys}}=\frac{p_i^{\text{phys}}}{\sqrt{\Enot\mst}}$. We then ought to write the Hamiltonian in terms of the (rescaled) canonical variables 
\begin{equation}
    \begin{split}
    Q^{\text{canon}}_{a} &= Q_{a}^{\text{phys}} + \Bp\e_{ab}P_b^{\text{phys}}\\
    P_{a}^{\text{canon}} &= P^{\text{phys}}_{a}
    \end{split}
\end{equation}
where $\Bp = \Enot \mst\Bmc $ is a dimensionless measure of the Berry curvature that scales as $\rs^{-\sfrac{3}{2}}$ when $\rs\to \infty$. From now on we will write $Q_i(P_i)$ for $Q_{i}^{\text{canon}}(P_{i}^{\text{canon}})$. 

A pair of valleys related by inversion symmetry has the same mass tensor but opposite Berry curvature. To describe the latter, we introduce an Ising variable $\mu = \pm 1 $ that we identify with the two possible valleys polarization within the pair so that we can write the Berry curvature of the pair of valleys as $ \bar{\Bmc} =\mu\abs{\bar{\Bmc}}$. 

The dimensionless effective Hamiltonian is
\begin{equation}\label{eq:EffBerryCurvature}
\begin{split}
    h_{\text{eff}}&=\frac{H_{\text{eff}}}{ \Enot}= h_{0}+\abs{\bar{\Bmc}}h_{1} + \abs{\bar{\Bmc}}^2h_{2};\\
      h_{0}&= \sum_{i} \frac{1}{2}P_{i}^2 +\sum_{i, j} \frac{1}{2} Q_iK_{ij}Q_j;\\
       h_{1} & = -\sum_{ij} \frac{\mu_i(\e P)_{i}K_{ij}Q_j+ \mu_jQ_jK_{ij}(\e P)_{i}}{2};\\ 
        h_{2} & = \sum_{ij} \frac{\mu_i\mu_j(\e P)_{i}K_{ij}(\e P)_{i}}{2};
\end{split}
\end{equation}
here the sums of $i$ and $j$ are over the sites of the triangular lattice. Note that the pseudo-spin pattern enters Eq.~\ref{eq:EffBerryCurvature} via the sign of the Berry curvature ($\mu_i$). {$\varepsilon P$ denotes the product of the Levi-Civita tensor and the vector $P$.} 

Recall that we are interested in the large $\rs$ limit. Therefore, we can treat $h_1$ and $h_2$ via degenerate perturbation theory over the ground state manifold of $h_0$.
The extensive degeneracy stems from the $\mu_i$ variables that do not enter $h_0$.

The leading effect of the Berry curvature to the energy is of order $\rs^{-3}$ and corresponds to the expectation value of $\Enot\abs{\bar{\Bmc}}{h_{1}}$ on the ground state of $h_0$. It turns out that $\expval{h_1} = 0 $ so that the leading effect of the Berry curvature on the energy comes with a prefactor of $\Enot \abs{\bar\Bmc}^2\propto\rs^{-9/2}$ which is parametrically smaller than the effect of trigonal warping that is generically present. Therefore, we will not calculate these higher order corrections of the Berry curvature to the energy.

\section{Extensions}\label{sec:Extensions}

We now consider two extensions of our methods to slightly more complicated systems:

\subsection{External moir\'{e} potential}

\def\kMo{\kappa_{\moire}}
{While in typical semiconductors, commensurability effects with the underlying lattice are negligible, if the period is extended - as is the case in moir\'{e} materials - such effects can be significant.}
Consider adding an external 
{periodic potential that 
is} commensurate with the Wigner crystal. 
Still keeping terms that to quadratic order in displacements about the classical ground-state, this corresponds to adding $\frac{1}{2}K_{\moire} \hat{x}_{i}^2$ at every site of the WC. The elastic matrix gets shifted by a diagonal matrix
\[
K_{\moire}(\qbs) = K(\qbs) + \kMo \Id; 
\]
We will treat $\kMo = \frac{K_{\moire}}{\Enot^2 \mst}$ as an additional parameter.
%
In the case of isotropic mass tensors, the (normalized) phonon energies are  
\[
\w_{\moire,A}(\qbs) = \sqrt{\w_{A}^2(\qbs)+ \kMo }, \quad A=1,2.
\]
{ In particular, as expected, this gaps the acoustic modes.

 We have redone the above calculations of the preferred orbital ordering in the presence of a commensurability potential for a subset of the various cases considered above.  
For instance, over the range of parameters we have explored, for the case of a tetragonal symmetry considered Sec.~\ref{sec:SmallMassAniPerturbative} in the limit of small mass anistropy $(\lambda\ll 1)$, 
we find that the pseudo-spin configuration with least energy is still the period-two stripe. 

We have also repeated the calculations of  Sec.~\ref{sec:cubicCorrection} in the presence of trigonal warping and no mass anisotropy, and again we find that commensurability effects do not change the character of the preferred orbital order.} 
To reach this conclusion, we evaluated the kernel of Eq.~\ref{eq:SecondOrderCorrection} with the modified dispersion for $\kMo \in \{1/2,1,2\}$ and, additionally, performed a large $\kMo$ expansion in App.~\ref{app:C2breaking}. In both regimes the configuration with least energy corresponds to $\ee^{3\ii \phi_i} = \pm \ii$.

{ Needless to say, the biggest effects of a moir\'{e} potential (which we do not explore here) is felt if the system is slightly incommensurate.  Here, rather than favoring a WC with an incommensurate period, the preferred state is a commensurate WC with a small concentration of interstitials or vacancies.  In a recent study \cite{kim2022interstitial}, it was shown that a small concentration of interstitials can have an outsized effect on the magnetic (spin) order of a WC.}

\subsection{3 dimensional Wigner crystals}

We can adapt our considerations to the 3-dimensional electron gases with multiple valleys. In the following, we go through the analysis and make emphasis on the distinctions with the 2-dimensional problem. 

At large $\rs$, Coulomb interactions dominate and the system minimizes the energy by forming a WC with body-centered cubic structure \cite{fuchs1935Wigner3D} with lattice vectors $[\frac{1}{2},\frac{1}{2},\frac{1}{2}], [\frac{1}{2},-\frac{1}{2},\frac{1}{2}], [\frac{1}{2},\frac{1}{2},-\frac{1}{2}]$. The leading correction in $1/\rs$ to the energy is again the zero-point phonon energy, which depends on the pseudo-spin pattern. If $W$ is the inverse mass tensor, we take $\mst= \frac{3}{\Tr[W]}$ as the mass scale (which is equal for every valley related by the PG action) and the mass anisotropy as
\begin{equation}
    \begin{split}
        \l 
        &= \sqrt{\frac{1}{6}\Tr[\left(\mst W-\id\right)^2]} \\
        &= \frac{\sqrt{w_1^2+w_2^2+w_3^2-w_1w_2-w_2w_3-w_1w_3}}{w_1+w_2+w_3}
    \end{split}
\end{equation}
where $w_i$ are the eigenvalues of $W$.

For small $\lambda$, we can derive an effective model for the valley pseudo-spin as in Sec.~\ref{sec:SmallMassAniPerturbative}. The pseudo-spins are generalized clock variables with states in a group $S$. The elements of $S$ correspond to symmetry operations in the PG of the host crystal that permute the valleys. For instance, consider a system with cubic symmetry and valleys at $[\pi,0,0], [0,\pi,0], [0,0,\pi]$. Then, $S=\ZZ_3$ corresponds to 3-fold rotation along the $[1,1,1]$ axis. 

The effective energy per electron is better understood in terms of quadrupolar moments $Q_i$. These are related to the inverse mass tensor of electron at site $i$ by $W_i = (\id + \lambda Q_i)/\mst$. At every site, we associate to $\gbf_i\in S$ the quadrupolar moment obtained by applying $\gbf$ to a reference $Q_{\WC}$. ($Q_{\WC}$ plays the role of $\q_{\WC}$ from Sec.~\ref{sec:AnisotropicMassTensor}.)

The leading correction in $\lambda$ to the energy is $\l^2e_2$, where
\begin{equation}\label{eq:energyClock3d}
   e_2 = -\frac{1}{\Nel}\sum_{\kbs} \tilde{Q}^{\dag}_{\kbs}\tilde{R}(\kbs)\tilde{Q}_{\kbs}^{}.
\end{equation}
The sum is over the 3-dimensional BZ. $\tilde{Q}_{\kbs}$ is the Fourier transform of $Q_i$ (seen as a 5 component vector) and $\tilde{R}(\kbs)$ is a symmetric matrix (the expression can be found in Sec.~\ref{app:WCin3d}) obtained from the phonon spectrum of the 3d WC with $\l=0$.

We have evaluated $\tilde{R}(\kbs)$ in a $12\times 12\times 12$ momentum grid and found that the largest eigenvalue over all $\kbs$ arises at $\Nbs = [\pi,\pi,0] $ (and symmetry-related momenta) and has $e_2\approx - 0.01283$. The corresponding eigenvector $(\tilde{Q}_{\kbs}=\d_{\kbs,\Nbs}\tilde{Q}^*)$ is interpreted in terms of the mass tensors as follows. There is pseudo-spin ordering with a two-site unit cell (the new lattice vectors are $[1,1,1]$, $[\frac{1}{2},-\frac{1}{2},-\frac{1}{2}]$, $[\frac{1}{2},-\frac{1}{2},\frac{1}{2}]$). Equivalently, the pseudo-spins are constant on the $x+y=k $ planes but oscillate in the $[1,1,0]$ direction. 
The principal axes of the inverse mass tensors $W$ (and therefore of $Q$) are along the $x-y, x+y, z$ directions of the WC. $Q_i$ has eigenvalues 
\begin{equation}
    \begin{split}
        q_{x+y}^{(A)}&= 0; q_{x-y}^{(A)} = +\sqrt{3}; q_{z}^{(A)} = -\sqrt{3}\\
    q_{x+y}^{(B)} &= 0; q_{x-y}^{(B)} = -\sqrt{3}; q_{z}^{(B)} = +\sqrt{3}\\
    \end{split}
\end{equation}
where $q_{d}^{(Y)}$ is the eigenvalue of $Q_i$ (as a matrix) along the $d$ direction of the electron on sublattice $Y$.


{As in the 2d case, the allowed $Q_i$ configurations are determined by the host semiconductor. If we restrict to configurations with two sites per unit cell, we can write $\tilde{Q}_{\kbs} = \cos(\gamma)\d_{\kbs, \zero} \tilde{Q}_{\Gamma} + \sin(\gamma)\d_{\kbs, \Qbs} \tilde{Q}_{\Qbs}$. Here $\gamma$ is an angle, $\tilde{Q}_{\Gamma}$ and $\tilde{Q}_{\Qbs}$ are (properly normalized) symmetric traceless matrices, and $\Qbs$ is a crystal momentum such that $2\Qbs$ is a reciprocal lattice vector. 
In a BCC crystal, there are 8 such momenta: $\Gamma$, $\Nbs$ (plus other 5 symmetry related momenta) and $\Hbs=[2\pi,0,0]$.}

For instance, let us go back to the example of a host crystal with cubic symmetry and valleys at $[\pi,0,0]$ and related momenta. We optimize over states with unit cells with two sites and find that the lowest energy configuration has $e_2\approx-0.01267$ and corresponds to a pseudo-spin ferromagnet with the longitudinal axes of mass tensor aligned with one of the nearest-neighbour directions (e.g. $[1,1,0]$). Note that the energy of this configuration is larger than the energy of the configuration we discussed above but that configuration is not allowed by the host crystal band structure.

\section{Possibilities for thermal and quantum melting}\label{sec:Melting}

Even for the case of a single, isotropic valley, the nature of the  transitions from the WC to the symmetric fluid phase is unsettled.  The presence of long-range Coulomb interactions preclude a single first-order transition \cite{Spivak2004PhasesBetween2DELnWC}, making it likely that the melting is always a multi-step transition.  For the thermal melting, a two step Halperin-Nelson scenario with an intermediate hexatic phase is certainly possible \cite{Halperin1978_2dMelting}. The nature of the quantum melting (as a function of decreasing $r_s$) is still less well understood.


In the WC in a host with $C_4$ symmetry, the pseudo-spin order breaks the $C_4$ down to $C_2$ (Sec.~\ref{sec:nv_2}).  It is thus plausible that the orientational order survives for a range of $T$ and/or $r_s$ beyond the point at which the translational order of the WC is melted.  This naturally leads to the expectation that there is a two-stage melting process, via a Kosterlitz-Thouless transition from a pseudo-spin ordered WC to an electronic Ising-nematic fluid, followed by an Ising transition to the isotropic fluid phase.
Other possiblities are that there could be an orbital ordering transition within the WC phase from the period two striped phase we found at large $r_s$ and $T=0$ to a uniform pseudo-spin ordered WC phase at smaller $r_s$ or larger $T$.  It is also possible to imagine the existence of a smectic phase intermediate between the WC and the nematic phases.

In the WC in a host with $C_6$ symmetry, similar considerations apply.  Here, the rotation symmetry in the pseudo-spin ordered WC is broken from $C_6$ to $C_2$, so the corresponding nematic phase would be an electronic three-state-Potts nematic fluid. Moreover, because the orbital order also breaks chiral symmetry, there is the possibility of vestigial chiral order once the WC is melted - either coexisting with nematic order or as a pure chiral fluid phase.

In the case of AlAs, there is experimental evidence that as a function of increasing $r_s$, there is a transition to a fully pseudo-spin polarized Ising-nematic fluid at a smaller value of $r_s$ than the critical value for WC (or more directly insulating state) formation\cite{Hossain2021SpontaneousValleyPolarization}.  It is probably not possible for there to be a continuous transition from such a state to the pseudo-spin stripe ordered WC state we found for large $r_s$.  However, as already mentioned, it is not impossible to imagine that there is a transition within the WC phase from  a stripe ordered WC phase (as in Fig.~\ref{fig:nv_2}) at large $r_s$ to a fully pseudo-spin polarized WC (as in Fig.~\ref{fig:nv_1}) for somewhat smaller $r_s$. {Another possibility is that there may be a small extrinsic energy splitting between valleys (e.g. from a small strain) 
which would tip the delicate energy balance in favor of the fully polarized phase.}

\section{Summary}\label{sec:Summary}
The ``pure'' 2DEG  is one of the paradigmatic problems in condensed matter physics. Moreover, the accuracy of the  effective mass approximation in many lightly doped semiconductors permits this problem to be experimentally realized with high fidelity in carefully constructed semiconductor devices. Experimentally observable signatures of a WC phase at large $r_s$ 
have been studied  in devices of ever increasing quality \cite{PhysRevB.61.10905, PhysRevLett.65.2189, PhysRevLett.79.1353, PhysRevLett.60.2765, PhysRevB.38.7881, PhysRevLett.82.1744}.  However, even in the context of the effective mass approximation, there are typically some additional features of the 2DEG in real devices that are both important for interpreting experiments, and also conceptually interesting as extensions of the basic theory.  Specifically, depending on the symmetry and details of the band-structure of the semiconductor in question, the effective mass tensor can be anisotropic, there can be additional pseudo-spin degrees of freedom associated with distinct valleys of the underlying band-structure, and there can be forms of symmetry breaking that rely on corrections to the effective mass approximation (for example, trigonal warping).  

Here we have analyzed several robust effects of this additional solid-state richness on the nature of the WC phase at large $r_s$.  Notice that despite pertaining to the crystalline phase, these are all quantum effects {(forms of ``order by disorder'')} in that they all enter through the form of the effective electron kinetic energy operator.  Some of our salient results are:  1) When there is a valley pseudo-spin degree of freedom ($n_v>1$), pseudo-spin ordering always occurs at a temperature scale that is smaller than the classical ordering scale only by a factor of $r_s^{-1/2}$ or $r_s^{-2}$. 
2)  In a tetragonal semiconductor, with two valleys as shown in Fig. \ref{fig:valleys}-{\color{blue} a}, the valley ordered state is the period 2 striped {pseudo-spin antiferromagnet} state shown in Fig.~\ref{fig:nv_2} that also breaks rotational symmetry down to $C_2$ - i.e. it has a nematic component. {This case is potentially relevant to the insulating state of AlAs 2DEGS~\cite{hossain2020observation}.} 3)  For a hexagonal semiconductor with three valleys that transform trivially under time-reversal, as shown in Fig.~\ref{fig:valleys}-{\color{blue} b}, the ordered state, shown in Fig.~\ref{fig:nv_3}, is again a period 2 stripe {pseudo-spin antiferromagnet} state, but one with both nematic and chiral components.  4)  For a hexagonal semiconductor with two valleys that transform into one another under time-reversal, as shown in Fig. \ref{fig:valleys}-{\color{blue} c}, the ground-state is the orbital (valley) ferromagnet shown in Fig.~\ref{fig:trigonalReal}, which is both chiral and breaks time-reversal symmetry. {We thus expect this state to have orbital loop-current ordering. The same valley-polarized state is obtained in the case of a moir\'{e} potential with a commensurate density of electrons. Such a state may be realized in transition metal dichalcogenide heterobilayers, where generalized Wigner crystal states were recently discovered at commensurate fillings~\cite{xu2020correlated,regan2020mott,huang2021correlated}.} 

Long though this paper may be, our analysis is not exhaustive.  There are a variety of  other possible valley geometries, for example the one shown in Fig. \ref{fig:valleys}-{\color{blue} d}, that could give rise to still other forms of orbitally ordered crystalline states - but these can be straightforwardly analyzed by the methods introduced here. Since the valleys transform into one another, in all cases there is strong pseudo-spin orbit coupling, i.e. there typically do not exist even approximate symmetries that operate on the pseudo-spins while leaving the spatial coordinates invariant.  (Indeed, it is worth noting  that even if the effective mass tensor is isotropic, corrections to the effective mass approximation always break rotational symmetry, so the WC is always orientationally pinned with respect to the symmetry axes of the underlying semiconductor.)  If there is strong spin-orbit coupling in the underlying semiconductor, this can lead to strong coupling between the electron spin and pseudo-spin in the effective theory.  This, in turn, could lead to interesting consequences for spin ordering in the WC that we have not yet explored.

Finally, we note that there are presumably many possible ways that the orbital order we have established in the WC phase could impress itself on the proximate quantum or thermal phases reached by melting or partially melting the WC.  We have speculated a bit about this in Sec.~\ref{sec:Melting}, but this largely remains unexplored territory. 

\section*{Acknowledgements}

We acknowledge useful discussions with Akshat Pandey and with Ilya Esterlis who also provided helpful comments on the manuscript. This work was supported in part by the National Science Foundation under Grant DMR-2000987 at Stanford (SAK and VC). EB was supported by the European Research Council (ERC) under grant HQMAT (Grant Agreement No. 817799), the Israel-US Binational Science Foundation (BSF), and the Minerva Foundation. This work was performed in part at Aspen Center for Physics, which is supported by National Science Foundation grant PHY-1607611.

\bibliography{References}

\appendix
\section{Elastic matrix}\label{app:ElasticMatrixAtK}
The elastic matrix can be written as a sum of two parts (set $e^2=\mst a_c^{1/2}$ in the expressions of Ref.~\cite{Bonsall1977WignerCrystal2D})
\begin{equation}
    \begin{split}
        K(\kbs)
        &=  \Bmc^<(\kbs)+\Bmc^{>}(\kbs)
    \end{split}
\end{equation}
where each term is defined as
\begin{align}
    {\Bmc^{<}_{ab}(\kbf)}&=\left[ \sum_{\Gbs\in \L^\vee } \!\left(\Fmc_{ab}(\Gbs+\kbs)-\Fmc_{ab}(\Gbs)\right)\right]\\
    {\Bmc^{>}_{ab}(\kbf)}&=\left[ \sum_{\Rbs\in \L} \!2\sin[2](\kbs\cdot \Rbs/2)\Emc_{ab}(\Rbs)\right]\\
    \Fmc_{ab}(\qbs) &= (aq)_a (aq)_b \frac{\sqrt{\WBZ}}{q}\Erfc\left(\sqrt{\frac{\pi q^2}{\WBZ}}\right)\\
    \Emc_{ab}(\Rbs) &=a^2\pdv{}{R_{a}}{R_{b}} \frac{\sqrt{\W}}{R}\Erfc\left(\sqrt{\frac{\pi R^2}{\nu}}\right)
\end{align}
with $\nu=\frac{\sqrt{3}}{2}=(\nel a^2)^{-1}$ and $\WBZ = (2\p)^2/\nu $. We denote the WC lattice without the origin by $\L_{\times}$ and the reciprocal lattice without the origin by $\L_{\times}^{\vee}$. The sums converge fast because they are exponentially suppressed in the distance from the center of the respective lattice. 

Below we calculate the dispersion relation around high-symmetry points. These have interesting features that show up in the density of states.
\subsection{Dispersion relations}

\paragraph{$\kbs=\GGbs+\qbs$:} the elastic matrix is 
\begin{equation}
    \begin{split}
        K(\qbs)_{ab} 
        &= \frac{\xi_0}{\xi}\xi_a\xi_b+ \k^{\G}_{abcd}\xi_c\xi_d+ \mmrm{O}(\xi^3)
    \end{split}
\end{equation}
where $\xi= aq$, $\xi_{0}=\kWC a$ and 
\begin{equation}
    \k_{abcd}^{\G}= \k_{\parallel}^{\G}\frac{\d_{ac}\d_{bd}+\d_{ad}\d_{bc}}{2}+\k_{}^{\G}\frac{\eps_{ac}\eps_{bd}+\eps_{ad}\eps_{bc}}{2}.
\end{equation}
To leading order in $q$, the eigenvalues are 
\begin{equation}
    \begin{split}
        K^{+}(\qbs) 
        &= \xi_0 \xi+\k_{\parallel} \xi^2+\mmrm{O}(\xi^3)\\
        K^{-}(\qbs) 
        &= \k^\Gamma \xi^2+\mmrm{O}(\xi^3)
    \end{split}
\end{equation}
so to leading order
\begin{equation}
    \det(K(\qbs)) = \k^{\G}\xi_0 \xi^3.
\end{equation}
This means that even in the presence of anisotropic mass tensors, there will be always at least one gapless mode at $\qbs=0$. From our numerics, we find $\kappa_\parallel^{\Gamma} \approx -1.2253$ and $\kappa^{\Gamma} \approx +0.24506$. 

\paragraph{$\kbs=\Mbs_2+\qbs$:} we have an inverted (quadratic) band  and a saddle point. The Little group is $D_2$ instead of the full $D_{6}$. To quadratic order in $q$, the elastic matrix reads
\begin{widetext}
\begin{equation}
    \begin{split}
       \left( K(\Mbs_2+\qbs) \right)_{ab} &= k^{M_2}_{ab} + \kappa^{M_2}_{abcd}\xi_c\xi_d  \\
        k^{M_2} &=
\begin{pmatrix}
1.42218 & 0 \\
0& 11.451 
\end{pmatrix} \\
\kappa^{M_2} 
&= \left(
\begin{array}{cccc}
 1.58534 & 0 & 0 & -0.251287 \\
 0 & -1.13164 & -1.13164 & 0 \\
 0 & -1.13164 & -1.13164 & 0 \\
 -1.35849 & 0 & 0 & -0.629068 \\
\end{array}
\right)
    \end{split}
\end{equation}
\end{widetext}
where the indices of $\kappa^{M_2}$ are ordered as $\{xx,xy,yx,yy\}$. 

The eigenvalues to quadratic order in $q$ are
\begin{equation*}
    \begin{split}
        K^{-}(M_2+\qbs) = k^{M_2}_{11} +  \kappa_{1;1}^{M_2} \xi_x^2+\kappa_{1;4}^{M_2}\xi_y^2 \\
        K^{+}(M_2+\qbs) = k^{M_2}_{22} +  \kappa_{4;1}^{M_2} \xi_x^2+\kappa_{4;4}^{M_2}\xi_y^2 
    \end{split}
\end{equation*}
note that $ K^{-}$ is a saddle point while $K^+$ is a band top. 

\paragraph{ $\kbs=\pm\Kbs+\qbs$}, the little group is $C_3$ 

\begin{equation}
    \begin{split}
        K(\Kbs+\qbs) 
        &= k_{K} \t^0  + \xi^2\left(\k_0^K \t^0 + \k^K_1\t^3_{\q}\right) \\
\t^3_\q = &(\t^3\cos(2\q) +\t^1\sin(2\q))   \\
k_K &= 6.683 \\
\k_0^K &= -0.1794\\
\k_1^K &= -0.2691 \\
\frac{\k_0^K}{\k_1^K} &= \frac{2}{3} \text{, up to numerical precision}
    \end{split}
\end{equation}
where $\q$ is defined by $\qbs=(q\cos(\q),q\sin(\q))$. The dispersion relations are 
\begin{equation}
    \begin{split}
        K^{\pm}(\Kbs+\qbs) &=  k_{K} + \xi^2\left(\k_0^K \mp \k^K_1\right) 
    \end{split}
\end{equation}

\twocolumngrid
\section{Details about the Variational Approach}\label{app:VariationalApproach} 

Our trial states are of the form $\ket{\Psi}=\ket{\Psi_{v}(\{\a_{i}\})}\otimes \ket{\Psi_{\text{ph}}(\{\a_{i}\})}$ where $\ket{\Psi_{v}(\{\a_{i}\})}=\bigotimes_{i}\ket{\a_{i}}_{i}$ is a product state and $\ket{\Psi_{\text{ph}}(\{\a_{i}\})}$ is the ground state of the phonon Hamtiltonian with $\hat{W}_{ab;i}\rightarrow \mel{\a_i}{\hat{W}_{ab;i}}{\a_i} =:W_{ab;i}$. We consider periodic states $\ket{\Psi_{v}(\{\a_{i}\})}$ with possibly more than one (WC) lattice site per unit cell. We specify the larger unit cells by new lattice vectors $\bbs_{1}$ and $\bbs_{2}$ which must be integer combinations of the WC lattice vectors $\abss_1=a(1,0)$ and $\abss_2=\tfrac{a}{2}(1,\sqrt{3})$. In addition to the valley polarization patter, we also need to fix the overall orientation of the WC lattice, or equivalently, an orientation of the mass tensors with respect to the axes of the WC. We denote this angle by $\Qwc$. 

We define the Fourier transform (FT) in terms of the WC lattice and interpret the larger unit cell as folding the BZ to a smaller one. Our FT is taken to be
\begin{equation}
    \begin{split}
        Q_{\kbs;a} &=\frac{1}{\sqrt{N_{\text{el}}}}\sum_{i } \ee^{\ii \kbs\cdot \Rbs_i} Q_{i;a},\\
    P_{\kbs;a} &= \frac{1}{\sqrt{N_{\text{el}}}}\sum_{i } \ee^{\ii \kbs\cdot \Rbs_i} P_{i;a},
    \end{split}
\end{equation}
and the commutation relations read
\begin{equation}
\begin{split}
        [Q^{}_{\kbs;a},P^{\dag}_{\kbs';b}] 
        &=  \frac{\ii \d_{ab}}{N_{\text{el}}} \sum_{i} \ee^{\ii (\kbs-\kbs')\cdot \Rbs_i}\\
        &= \ii\d_{ab}\sum_{\Gbs}\d_{\kbs+\Gbs,\kbs'}
\end{split}
\end{equation}
where the last sum is over reciprocal lattice vectors. Note that $(Q_{\kbs})^{\dag}=(Q_{-\kbs})$ and $(P_{\kbs})^{\dag}=(P_{-\kbs})$ and $\kbs$ is understood as a momentum on the first Brillouin zone of the WC ($\BZWC$).

Next we plug in the definitions of the FT into the Hamiltonian in Eq.~\ref{eq:HamiltonianPhonon}. The terms that depend of the position operators reduce to
\begin{equation}
    \frac{H_{\hat{r}}}{N_{\text{el}}} =\int_{\kbs \in \text{BZ}}\frac{\dd \kbs^2}{\WBZ} \sum_{ab}\frac{1}{2}Q_{\kbs;a}^{\dagger}K(\kbs)_{ab}Q_{\kbs;b}^{}.
\end{equation}
where $\WBZ= \frac{4\p^2}{\nu}$ is the area of $\BZWC$ and $K(\kbs)$ is the elastic matrix which corresponds to the Fourier transform of the dipole-dipole interactions. See App.~\ref{app:ElasticMatrixAtK} for the explicit expressions.

The $\hat{p}$ part of $H$ is off-diagonal in momentum space. The larger unit cell  spanned by $\bbs_1$ and $\bbs_2$ has an area that is an integer multiple of the original unit cell: $\Wwc^{'} = B\times  \Wwc^{}$. When we take the FT of $W_{ab;i}$ we will generally find weight at $B$ different momenta in the $\BZWC$ that are closed under addition modulo reciprocal lattice vectors. We denote this set by $\Lmc=\{\Lbs_1,\dots,\Lbs_B\}$.
Then 
\begin{equation}\label{eq:Hp11}
    \begin{split}
        \frac{H_{\hat{p}}}{N_{\text{el}}} &=\int_{\kbs \in \text{BZ}}\frac{\dd \kbs^2}{\WBZ} \sum_{ab}\sum_{\Lbs\in \Lmc}\frac{\tilde{W}(\Lbs)_{ab}}{2B}P_{\kbs+\Lbs;a}^{\dagger}P_{\kbs;b}^{};\\
        &=\int_{\kbs \in \text{BZ}'}\frac{\dd \kbs^2}{\WBZ} \sum_{\a\b}\Wmc_{\a\b}\frac{\Pmc_{\kbs;\a}^{\dagger}\Pmc_{\kbs;\b}^{}}{2};
    \end{split}
\end{equation}
where $\a=(a,l)$ is a composite index with $a= x,y$ and $l=1,\dots,B$. The new variables are $\Qmc_{\kbs;(a,l)}:=Q_{\kbs+\Lbs_l;a}$, $\Pmc_{\kbs;(a,l)}:=P_{\kbs+\Lbs_l;a}$,  $\Wmc_{(a,l)(b,l')}=\tilde{W}(\Lbs_{l}-\Lbs_{l'})_{ab}$ and $\Kmc_{(a,l)(b,l')}(\kbs)=K(\kbs+\Lbs_{l})_{ab}\d_{ll'}$. Using the fact that $\Wmc$ is positive-definite, we write $\Wmc=\Vmc^2$ with $\Vmc$ positive-definite. Then we change variables to $\tilde{\Pmc}=\Vmc\Pmc$ and $\tilde{\Qmc}=\frac{1}{\Vmc}\Qmc$. The frequency matrix is $\W^2_{\kbs}=\Vmc\Kmc(\kbs) \Vmc$ and the energy becomes
\begin{equation}\label{eq:EphononZeroPtApp}
    \begin{split}
        \eph=\frac{E_{\text{ph}}}{N_{\text{el}}}
        &=\frac{1}{2}\int_{\kbs \in \text{BZ}'}\frac{\dd \kbs^2}{\WBZ}\Tr[\sqrt{\W_{\kbs}^2}]. 
    \end{split}
\end{equation}
\subsection{Overview of calculation}

Our calculations of the variational states were done as follows
\begin{enumerate}
    \item We chose a super-cell ($\bbs_1$ and $\bbs_2$) with a pattern of valley polarization.
    \item We numerically evaluate the zero-point energy in Eq.~\ref{eq:EphononZeroPt}, $\eph$, for several values of $\eta$ and an evenly spaced grid of $\Qwc$. The numerical evaluation used an evenly spaced momentum grid of $24\times 24$ momentum points that respect the PG of the triangular lattice to approximate the integral over the BZ. 
    \item For a given $\eta$, we determine the optimal $\Qwc^\star(\eta)$ that minimizes $\eph$. We always find that for a given valley-polarization pattern, the value $\Qwc^\star(\eta)$ does not depend on $\eta$.
    \item For each $\eta$, we compare $\eph$ for the valley patterns (at their optimal $\Qwc$) and find the ones with the lowest $\eph$. We always find that the order of $\eph$ as a function of valley patterns is independent of $\eta>0$.
    \item To confirm the lowest energy trial state, we recalculate $\eph$ for the two configurations with lowest $\eph$ for a finer $\eta$ grid.
\end{enumerate}
\section{Perturbative calculations around the isotropic electron gas}
\subsection{Correlation functions for isotropic case}
We record some correlation functions for the isotropic case. We start with 
\begin{equation}
    \begin{split}
        h &= \sum_{i}\frac{P_i^2}{2}+\sum_{ij}\frac{Q_iK_{ij}Q_j}{2} \\
        &= \sum_{\kbs} \frac{P^{\dag}_{\kbs}P^{}_{\kbs} +Q^{\dag}_{\kbs}K(\kbs)Q^{}_{\kbs}}{2}
    \end{split}
\end{equation}
We then introduce the harmonic modes as follows. For $K(\kbs)v_{\kbs;A} = \w_{\kbs A}^2 v_{\kbs;A}$ where $A$ is the "band index". Then we introduce oscillator variables as follows 
\begin{equation}
    \begin{split}
        Q_{\kbs}  &= \sum_A\sqrt{\frac{1}{2\w_{\kbs;A}}}v_{\kbs;A}\left(\a^{\dag}_{\kbs;A}+\a^{\,}_{-\kbs;A}\right), \\
        P_{\kbs}  &= \sum_A\ii\sqrt{\frac{\w_{\kbs;A}}{2}}v_{\kbs;A}\left(\a^{\dag}_{\kbs;A}-\a^{\,}_{-\kbs;A}\right).
    \end{split}
\end{equation}

Then ($\expval{O}\equiv \Tr[O \ee^{-\beta h}]/\Tr[\ee^{-\beta h}]$)
\begin{equation}
    \begin{split}
        \expval{P_{i;a}(\tau)P_{j;b}} &=  \frac{1}{\Nel}\sum_{\qbs }\ee^{\ii \qbs\cdot\Rbs_{ij}}\expval{P^{\dag}_{\qbs;a }(\tau)P_{\qbs;b}} \\
    \expval{P^{\dag}_{\qbs;a }(\tau)P_{\qbs;b}} &= \sum_{A}\frac{\w_{\kbs;A}}{2}v_{\kbs;A;a}v_{\kbs;A;b}G_b(\w_{\kbs;A},\t) \\
    G_b(\w,\t) & = \frac{\cosh(\w(\beta/2-\tau))}{\sinh(\frac{\beta \w}{2})}; 0\le \t \le \beta.
\end{split}
\end{equation}
\subsection{Mapping to effective clock model}\label{app:MapToClockModel}

The aim of this section is to obtain the mapping used in Sec.~\ref{sec:SmallMassAniPerturbative} to map the valley problem at small mass anisotropy ($\l\ll 1$) to an effective classical XY model with strong pinning potential. In this section, we are measuring energy in units of $\Enot$.

We first write the effective Hamiltonian in Eq.~\ref{eq:HamiltonianPhonon} as $h=h_0+\eps h_1$ with $\eps=\l/2$. $h_0$ is the Hamiltonian for the isotropic problem and $h_1$ is of the form 
\begin{align*}
    h_1 &= \sum_{i} P_i D_i P_i; \\
    D_i &= \overline{W_i}- 1 = \cos(2\q_{i})\t^3 + \sin(2\q_{i})\t^1,
\end{align*}
here $\q_{i}$ is the relative angle of the long axes of the Fermi surface of valley $\a_i$ with the horizontal axes. $\tau^i$ is $i$-th Pauli matrix.

Consider now the partition function of the valley pseudo-spins and the position degrees of freedom 
\[
Z = \sum_{\{\q_i\}} Z[\{\q_i\}] ; 
\]
where $Z[\{\q_i\}]$ is the partition function of the position degrees of freedom in the presence of a valley pattern $\{\q_i\}$. As $\eps$ is small, we can write 
\[
Z[\{\q_i\}] = Z_0 \times \exp(- \beta\var{F}[\{\q_i\}] )
\]
where $Z_0$ is the partition function of the isotropic problem $Z_0 =\Tr[\ee^{-\beta h_0}]$ and $\var{F}$ can be expanded in powers of $\eps$  so that $\var{F}[\{\q_i\}]=\sum_{n=0}^{\infty} \eps^n F_n[\{\q_i\}]$. The leading correction with $\{\q_i\}$ dependence is
\begin{equation}
    -\beta F_2= \frac{1}{2} \int^{\beta}_{0}\!\!\dd{\t} \int^{\t}_0\!\!\dd{\t'} \expval{h_1(\t)h_1(\t')}.
\end{equation}
The $\{v_i\}$ dependence is hidden in the definition of $h_1$.
We expand $h_1$ in terms of the momentum variables, evaluate the four point correlation functions and take the Fourier transform to arrive at
\begin{equation}\label{eq:freeEnergyApp}
    \begin{split}
       F_2 = -\frac{T }{2}\sum_{\kbs,\a,\a'}\d_{\kbs,\qbs+\qbs'} \Tr[\tilde{D}_{\kbs}\Pi_{\a}\tilde{D}_{\kbs}^{\dag}\Pi_{\a'}]\Gmc_{\beta}(\w_{\a},\w_{\a'})  
    \end{split}
\end{equation}
here $\a =(\qbs,A)$ and $A$ is the 'band' label of the phonon at momentum $\qbs$. $\Pi_{\a}=v_\a^{\top}v_{\a}$ is a projector where $v_{\a}=[\cos(\q_{\a}),\sin(\q_{\a})]^{\top}$ is the eigenvector of $K(\qbs_{\a})$ with eigenvalue $\w_{\a}^2$. $\tilde{D}_{\kbs} = \frac{1}{{\Nel}}\sum_{r}\ee^{\ii \qbs\cdot \Rbs_i} D_{i}$. $\Gmc_{\beta}(\w_{\a},\w_{\a'}) $ is given by
\[
\Gmc_{\beta}(\w_1,\w_2)= \frac{\sinhc(\frac{\beta(\w_1+\w_2)}{2})+\sinhc(\frac{\beta(\w_1-\w_2)}{2})}{\sinhc(\frac{\beta\w_1}{2})\sinhc(\frac{\beta\w_2}{2})},
\]
where $\sinhc(x)=\sinh(x)/x$. If we write $\tilde{D}_{\kbs}= \Phi_{2;\kbs}^{(1)} \t^3 + \Phi_{2;\kbs}^{(1)} \t^1 $, then $\Phi_{q;\kbs}^{(1)}$ and $\Phi_{q;\kbs}^{(2)}$ are the Fourier transform of $\cos(2\q_i)  $ and $\sin(2\q_{i})$, respectively. In terms of this variables, the correction to the free energy is 
\begin{equation}
     F_2=-\frac{1}{4}\sum_{\kbs} \Phi_{2;\kbs}^{\dag} \underline{K}(\kbs,\beta) \Phi_{2;\kbs}^{}
\end{equation}
where $\Phi_{2;\kbs}=[\Phi_{2;\kbs}^{(1)},\Phi_{2;\kbs}^{(2)}]^{\top}$. The kernel can be expanded as $\underline{K}= K_1\t^0 + K_2\t^3 + K_3\t^1$ with
\begin{equation}\label{eq:KernelsF2}
    \begin{split}
        K_{1}(\kbs,\beta) &= \frac{T}{\Nel}\sum_{\kbs,\a,\a'} \d_{\kbs,\qbs+\qbs'} \Gmc_{\beta}(\w_\a,\w_{\a'});\\
        K_{2}(\kbs,\beta) &= \frac{T}{\Nel}\sum_{\kbs,\a,\a'} \d_{\kbs,\qbs+\qbs'} \Gmc_{\beta}(\w_\a,\w_{\a'})\cos(2 (\q_\a+\q_{\a'}));\\
        K_{3}(\kbs,\beta) &= \frac{T}{\Nel}\sum_{\kbs,\a,\a'} \d_{\kbs,\qbs+\qbs'} \Gmc_{\beta}(\w_\a,\w_{\a'})\sin(2(\q_\a+\q_{\a'})).
    \end{split}
\end{equation}

The effective energy (at zero temperature) can be derived by using the relation $E=\pdv{\beta F}{\beta}\eval_{T=0}$. We use the identity $\pdv{\beta}\Gmc_{\beta}(\w_1,\w_2)\eval_{T=0} =\frac{\w_1\w_2}{\w_1+\w_2}  $ to write the leading contribution to the energy as 
\begin{equation}\label{eq:E2Clock}
     E_2=-\frac{1}{4}\sum_{\kbs} \Phi_{2;\kbs}^{\dag} \underline{K}(\kbs) \Phi_{2;\kbs}^{}
\end{equation}
where 
\begin{equation}\label{eq:KernelsE2}
    \begin{split}
    \underline{K}&=K_1 + K_2\t^3 + K_3 \t^1;\\
        K_{1}(\kbs) &= \frac{1}{\Nel}\sum_{\kbs,\a,\a'} \d_{\kbs,\qbs+\qbs'} \frac{\w_{\a}\w_{\a'}}{\w_{\a}+\w_{\a'}} ;\\
        K_{2}(\kbs) &= \frac{1}{\Nel}\sum_{\kbs,\a,\a'} \d_{\kbs,\qbs+\qbs'} \frac{\w_{\a}\w_{\a'}}{\w_{\a}+\w_{\a'}} \cos(2 (\q_\a+\q_{\a'}));\\
        K_{3}(\kbs) &= \frac{1}{\Nel}\sum_{\kbs,\a,\a'} \d_{\kbs,\qbs+\qbs'} \frac{\w_{\a}\w_{\a'}}{\w_{\a}+\w_{\a'}} \sin(2(\q_\a+\q_{\a'})); \\
      \tilde{J}(\kbs) &\equiv \frac{1}{16}\underline{K}(\kbs).
    \end{split}
\end{equation}
Eq.~\ref{eq:XYenergy} in the main text is obtained by diving Eq.~\ref{eq:E2Clock} by the number of electrons $(\Nel)$.

\begin{figure}[b]
        \centering
\subfloat[\label{fig:K123T=0}]{
  \centering
  \includegraphics[width=0.4\linewidth]{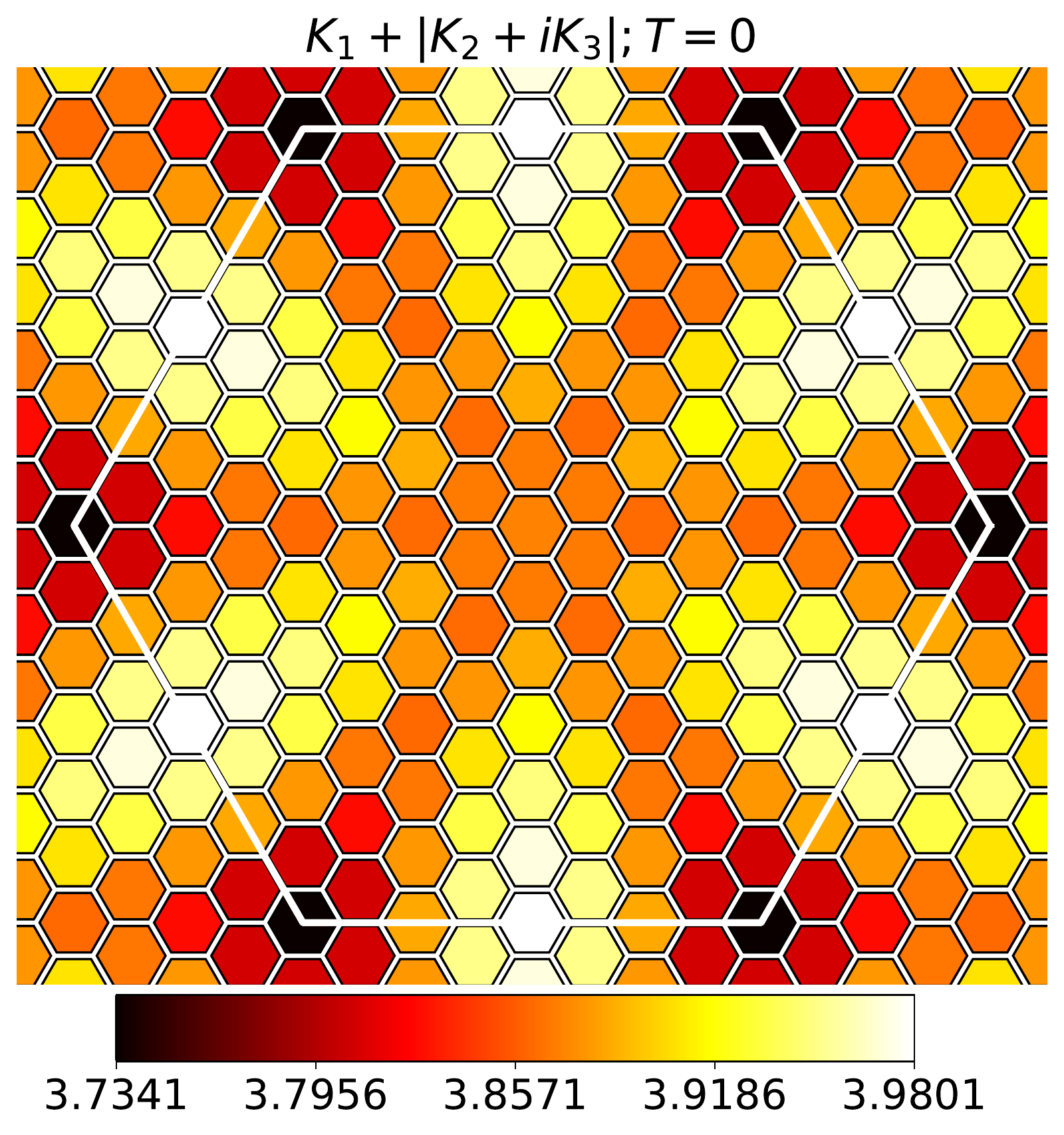}
}
\subfloat[\label{fig:K1T=0}]{
  \centering
  \includegraphics[width=0.4\linewidth]{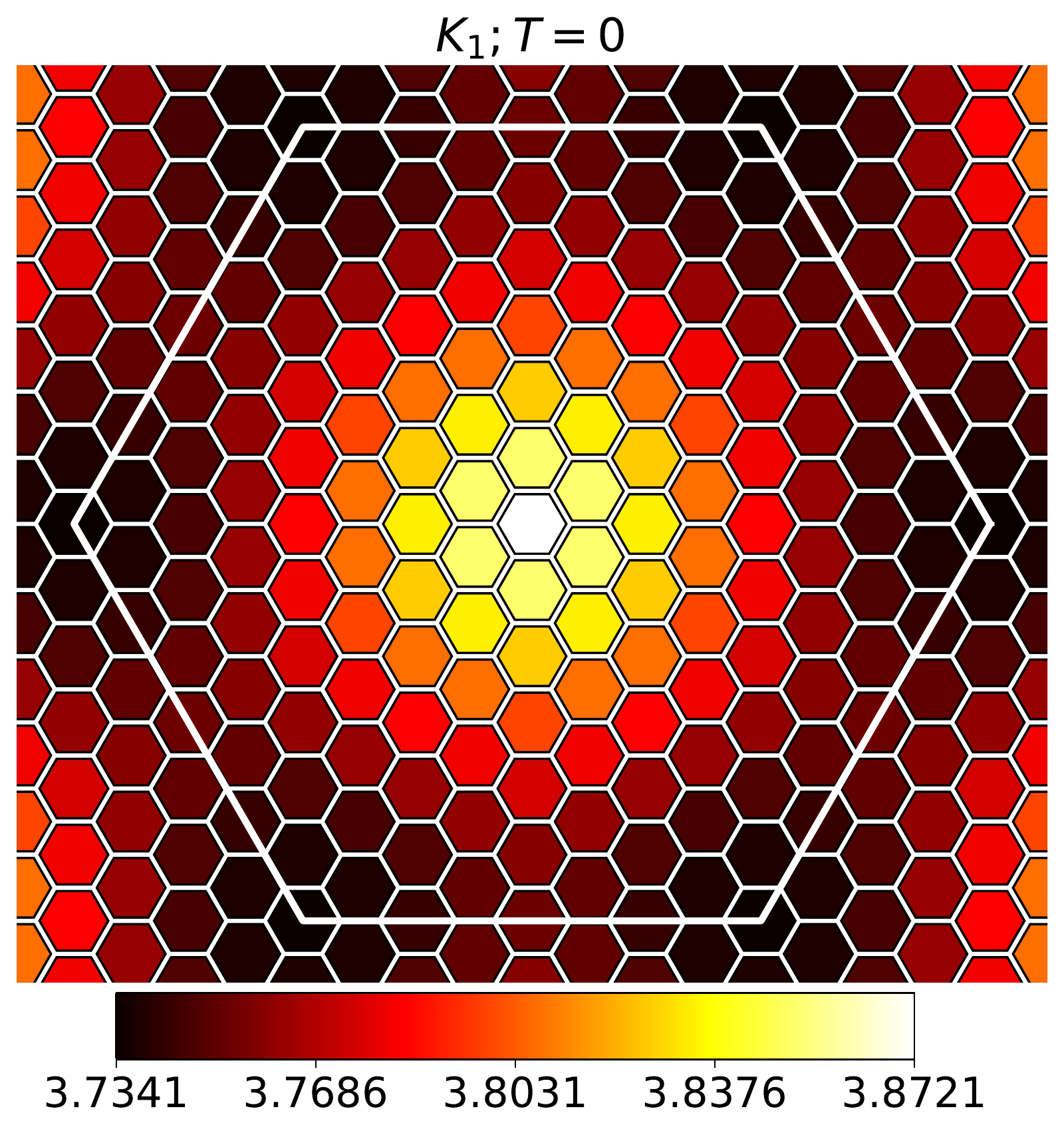}
}
\\
\subfloat[\label{fig:K23AbsT=0}]{
  \centering
  \includegraphics[width=0.4\linewidth]{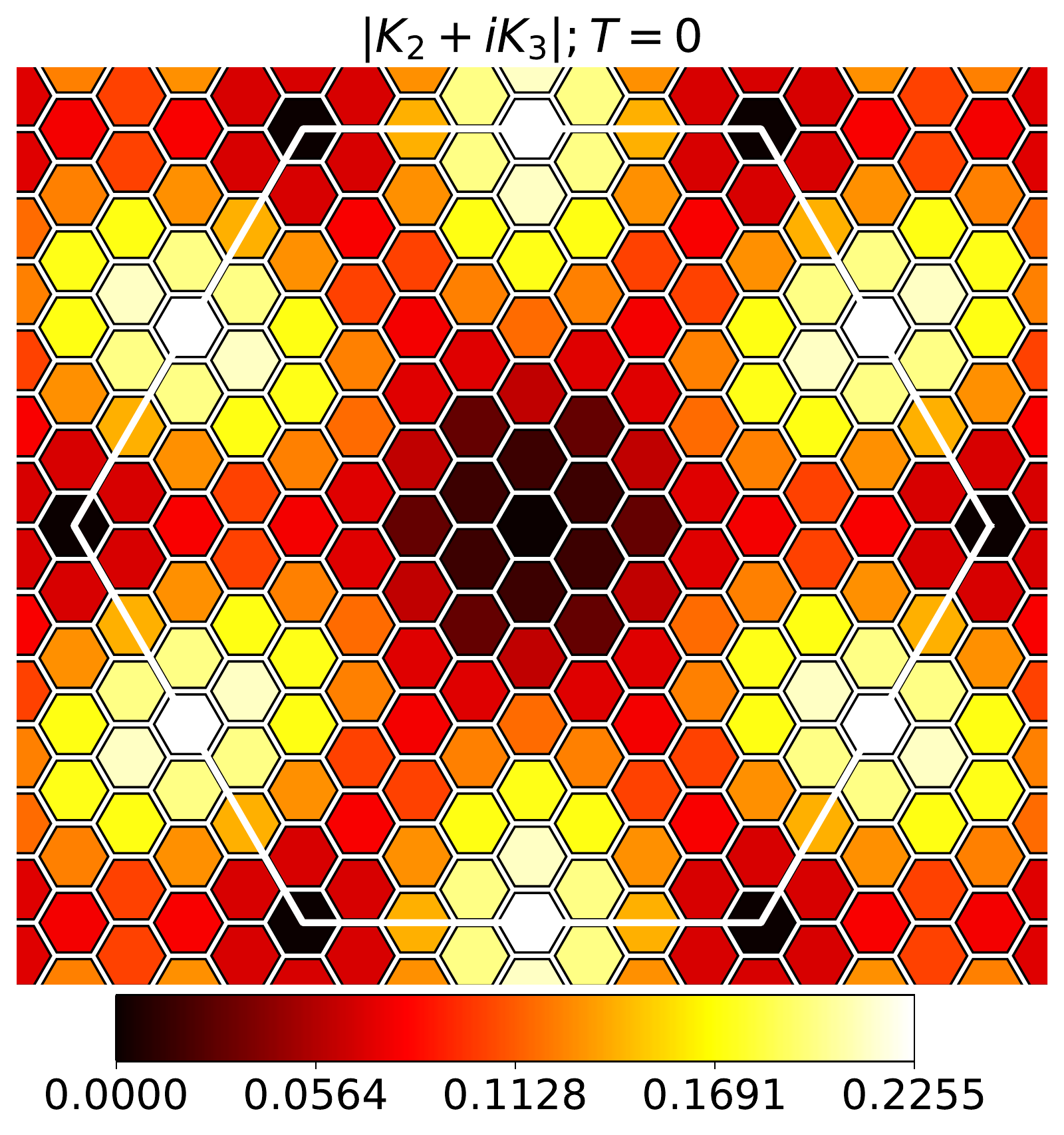}
}
\subfloat[\label{fig:K23ArgT=0}]{
  \centering
  \includegraphics[width=0.4\linewidth]{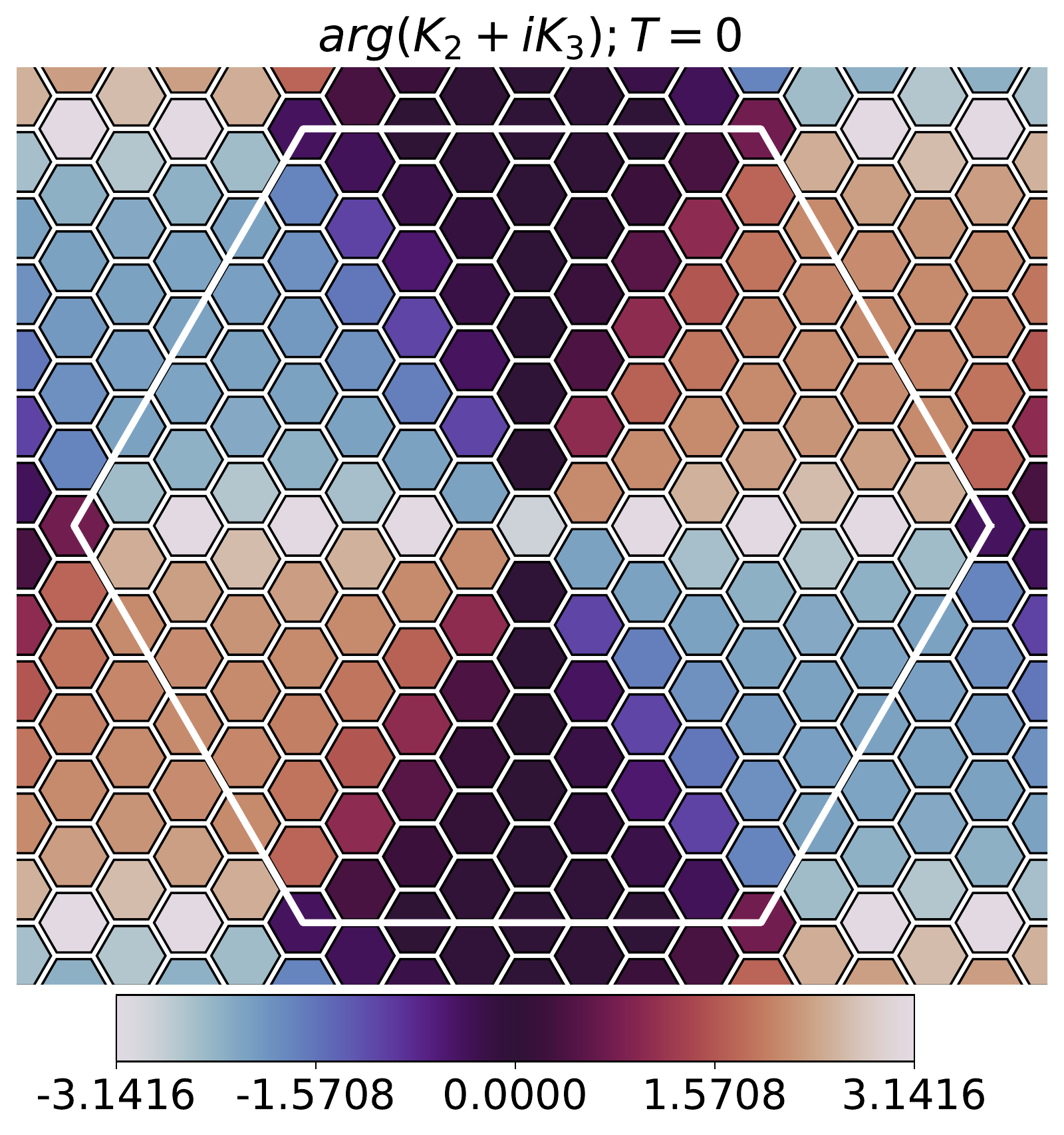}
}
\caption{\label{fig:KernelsE2} Plots of the components of the kernel in Eq.~\ref{eq:KernelsE2} used to evaluate the effective energy of the system.}
\end{figure}

We numerically evaluate $\underline{K(\kbs)}$ over a $12\times 12$ grid in momentum space. Fig.~\ref{fig:KernelsE2} shows each component as well as the largest eigenvalue ($K_1+\abs{K_2+\ii K_3}$) at each $\kbs$.

\subsection{Correction to the Energy from inversion pairs }\label{app:C2breaking}

We want to calculate the effect of trigonal warping on the degeneracy between two valleys related by inversion. As mentioned in Sec.~\ref{sec:InversionPairs}, the effective Hamiltonian (Eq.~\ref{eq:HTrigonalwarping}) is $h_{\text{eff}}= h_0 + \eps_3 h_3$ where $h_0$ is the phonon Hamiltonian in the absence of trigonal warping, $\eps_3\ll 1$ and 
\begin{equation}\label{eq:AppTrigonalHam}
    h_3 = \sum_{i} \ee^{+3\ii \phi_i}(P_{i;x}-\ii P_{i;y})^3+\ee^{-3\ii \phi_i}(P_{i;x}+\ii P_{i;y})^3
\end{equation}

We want to consider a perturbative expansion of the energy (per electron) with trigonal warping $e_{\text{ph}}^{\triangle}=\sum_{n=0}^{\infty}\eps_3^ne_n^{\triangle}$. $e_0^{\triangle}$ is the energy of the system in the absence of trigonal warping. We calculate $e_n^{\triangle}$ using many-body perturbation theory. 

First, we rewrite Eq.~\ref{eq:AppTrigonalHam} in momentum space 
\begin{equation}\label{eq:AppTrigonalHamK}
    h_3 = \frac{1}{\Nel^{1/2}}\sum_{\kbs_{1},\kbs_{2},\kbs_{3}} 
    \left[
    \Phi_{3;\qbs}^{+} \prod_{l=1}^{3} (\zeta^*\cdot P^{\dagger}_{\kbs_l}) + h.c.\right]
\end{equation}
here $\zeta=[1,\ii]$ and $\Phi_{3;\qbs}^{+}$ is the Fourier transform of $\ee^{3\ii \phi_i}$ with $\qbs=\kbs_1+\kbs_2+\kbs_3$.

Let's start with $e_1^{\triangle}$. This corresponds to evaluating matrix elements of $h_3/\Nel$ between unperturbed ground states. These states correspond to states with different center of mass momenta ( $P_{CM} = \frac{1}{\Nel}\sum_{i}P_i= \frac{1}{\sqrt{\Nel}}P_{\zero}$) and the respective ground state of the oscillators at the other crystal momenta. The only terms in Eq.~\ref{eq:AppTrigonalHamK} that contribute are the ones with an odd number of $\kbs_l=\zero$ and $\qbs=0$. If we write $\Phi_{3;\zero}^{+}=\abs{\Phi_{3;\zero}}\ee^{\ii \Q_{3;\zero}}$ and $P_{CM} = \abs{P_{CM}}[\cos(\q_{CM}), \sin(\q_{CM})]^{\top}$, then 
\begin{equation}
    \begin{split}
        e_1^{\triangle} 
        &= 2\abs{\Phi_{3;\zero}}\abs{P_{CM}}^3 \cos(3\q_{CM}-\Q_{3;\zero}) \\
        &+3\abs{\Phi_{3;\zero}}\abs{P_{CM}}\left[\ee^{\ii (\Q_{3;\zero}-\q_{CM})} \Amc^*+ c.c.\right]\\ 
        \Amc &= \frac{1}{2\Nel}\sum_{\kbs_{1},\kbs_{2}}  \expval{ (\zeta\cdot P^{}_{\kbs_1})  (\zeta\cdot P^{}_{\kbs_2})}
    \end{split}
\end{equation}
Let's assume that the ground state of $h_0$ preserves the $C_3$ symmetry of the WC so that $\Amc$ vanishes (to see this, under $C_3: \zeta\cdot P_{\kbs} \to \ee^{\frac{2\pi \ii}{3}}\zeta\cdot P_{C_3\kbs}$. Then, after relabeling the moment, we see that $\Amc \to \Amc \ee^{\frac{4\pi \ii}{3}}$ which implies that $\Amc=0$. ). We need to add a $\frac{u}{4}\abs{P}_{CM}^4$ to the energy ($u>0)$ to stabilize the system. The effective potential is then $e_0^{\triangle}+ \eps_3 e_1^{\triangle} + \frac{u}{4}\abs{P}_{CM}^4$ is minimized at $P_{CM}=0$ as long as $\eps_3$ is small because there is no linear term in $P_{CM}$.

Next, we evaluate $e_2^{\triangle}$ using perturbation theory and ignore the terms involving $P_{CM}$ as these terms are higher order in $\eps_3$ than the ones already present in $e_0^{\triangle}+ \eps_3 e_1^{\triangle} + \frac{u}{4}\abs{P}_{CM}^4$:
\begin{widetext}
\begin{equation}
    e_2^{\triangle}= -\frac{3!}{2^3\Nel^2}\sum_{\kbs\a_1,\a_2,\a_3}\d_{\kbs,\sum_{\a}\qbs_{\a}} \frac{\w_{\a_1}\w_{\a_2}\w_{\a_3}}{\w_{\a_1}+\w_{\a_2}+\w_{\a_3}} 
        \abs{\Phi_{3;\kbs}^{+}\ee^{-\ii \sum_{\a} \q_\a}+\Phi_{3;\kbs}^{-}\ee^{+\ii \sum_{\a} \q_\a}}^2
\end{equation}
\end{widetext}
As in Sec.~\ref{app:MapToClockModel}, we can expand the absolute value and write the final answer as 
\begin{equation}
   e_2^{\triangle}=  -\frac{3!}{2^4\Nel}\sum_{\kbs}{{\Phi}}_{3;\kbs}^{\dag} \underline{K}^{(3)}(\kbs) {\Phi}_{3;\kbs} 
\end{equation}
with ${\Phi}_{3;\kbs} $ the Fourier transform of $[\cos(3\q_i),\sin(3\q_i)]^{\top}$ and the components of kernel is
\begin{equation}\label{eq:KernelTRS}
    \begin{split}
    \underline{K}^{(3)} &= {K}^{(3)}_1 \t^0 + {K}^{(3)}_2 \t^3+ {K}^{(3)}_3 \t^1 \\
      K_{1}^{(3)}(\kbs)&= \frac{1}{\Nel^2}\sum_{\a} \d_{\kbs,\sum_{\a}\qbs_{\a}}   \frac{\w_{\a_1}\w_{\a_2}\w_{\a_3}}{\w_{\a_1}+\w_{\a_2}+\w_{\a_3}}; \\
      K_{2}^{(3)}(\kbs)&= \frac{1}{\Nel^2}\sum_{\a} \d_{\kbs,\sum_{\a}\qbs_{\a}}   \frac{\w_{\a_1}\w_{\a_2}\w_{\a_3}}{\w_{\a_1}+\w_{\a_2}+\w_{\a_3}} \cos(2\sum_{\a}\q_{\s});\\
      K_{3}^{(3)}(\kbs)&= \frac{1}{\Nel^2}\sum_{\a} \d_{\kbs,\sum_{\a}\qbs_{\a}}   \frac{\w_{\a_1}\w_{\a_2}\w_{\a_3}}{\w_{\a_1}+\w_{\a_2}+\w_{\a_3}} \sin(2\sum_{\a}\q_{\s});\\
      \tilde{J}_{3}(\kbs)&\equiv \frac{3!}{2^4} \underline{K}^{(3)}(\kbs).
    \end{split}
\end{equation}

\begin{figure}[h]
\subfloat[\label{fig:K123TRS}]{
  \centering
  \includegraphics[width=.4\linewidth]{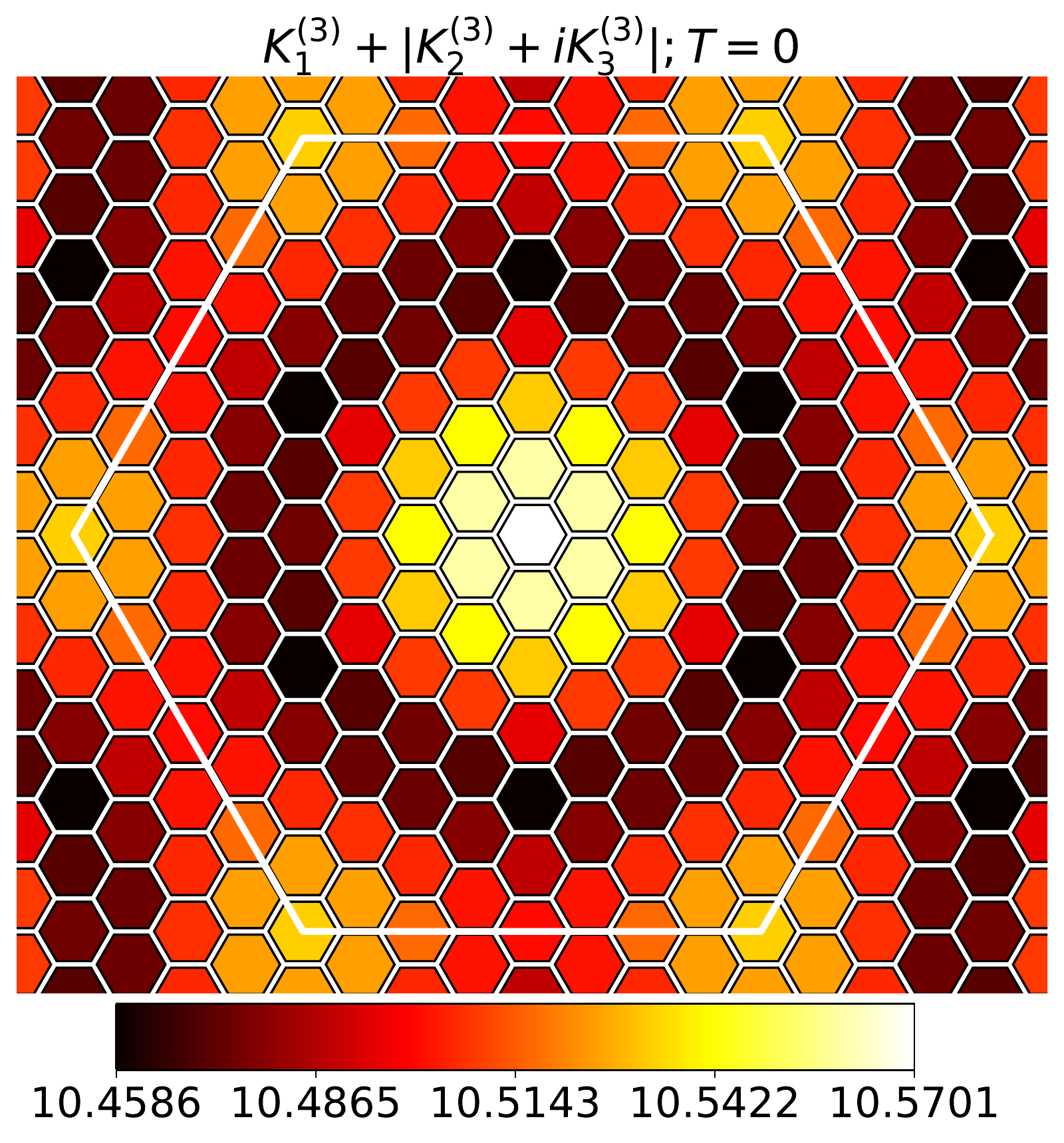}
}
\subfloat[\label{fig:K1TRS}]{
  \centering
  \includegraphics[width=.4\linewidth]{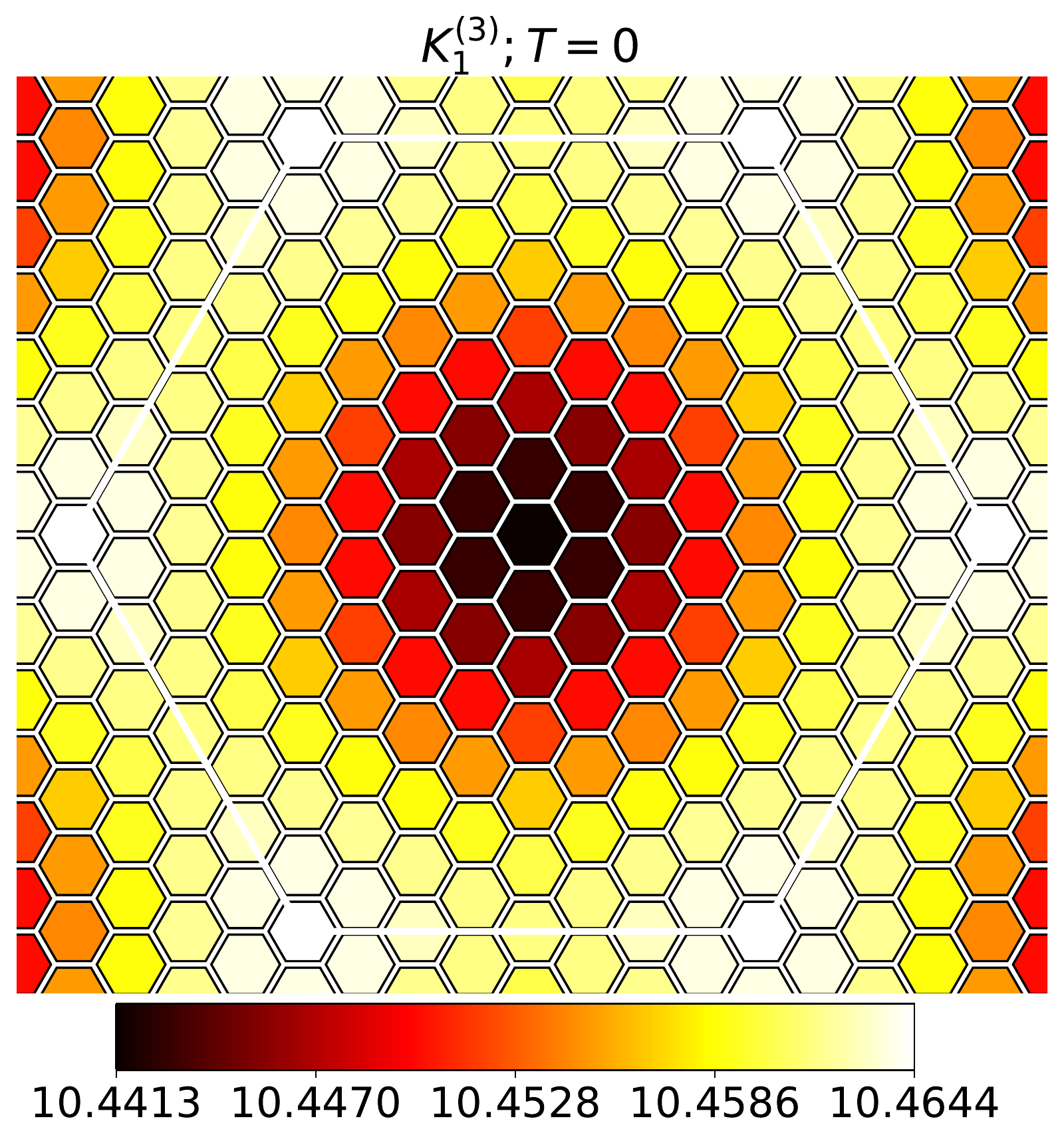}
}
\\
\subfloat[\label{fig:K23AbsTRS}]{
  \centering
  \includegraphics[width=.4\linewidth]{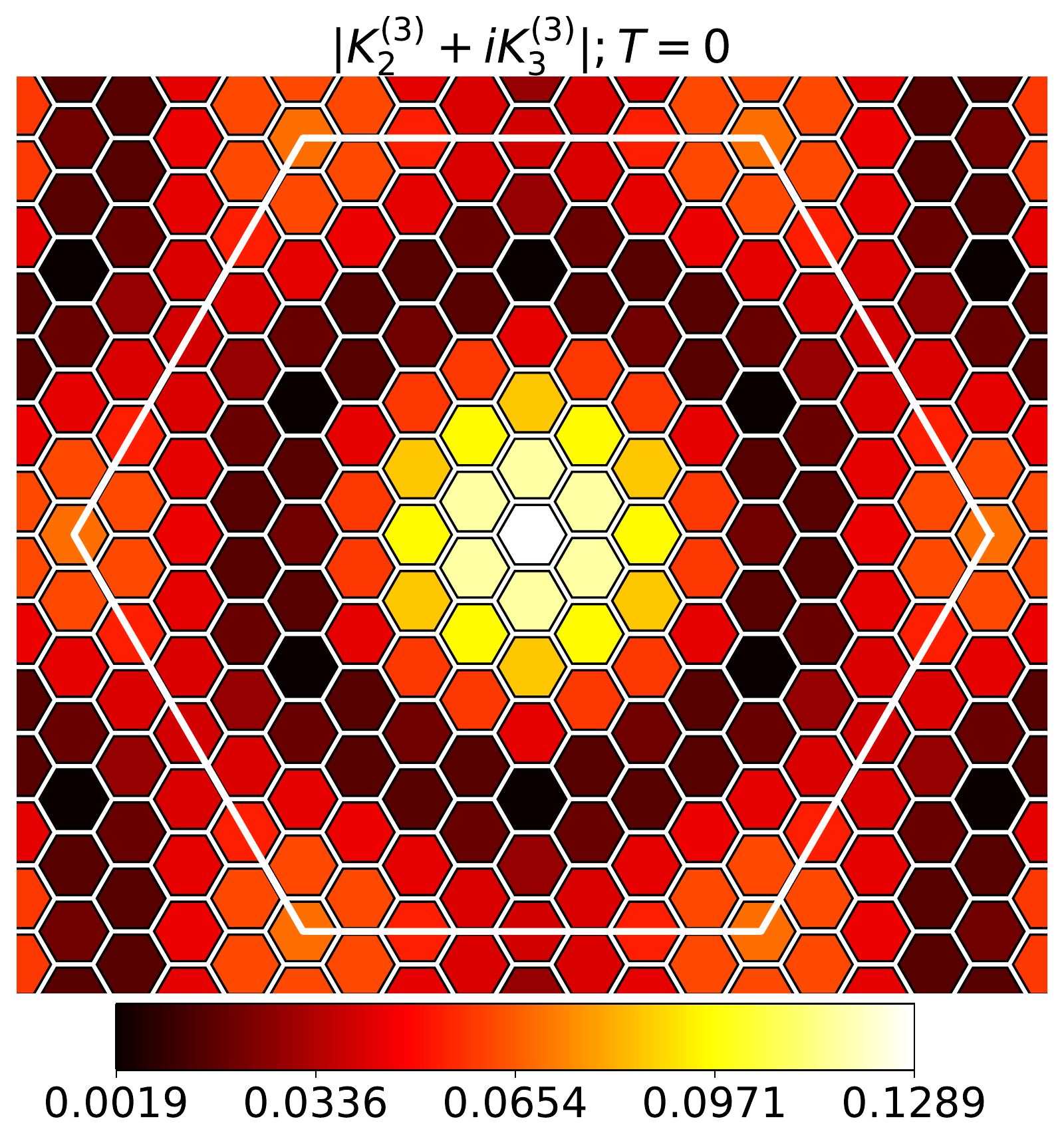}
}
\subfloat[\label{fig:K23ArgTRS}]{
  \centering
  \includegraphics[width=.4\linewidth]{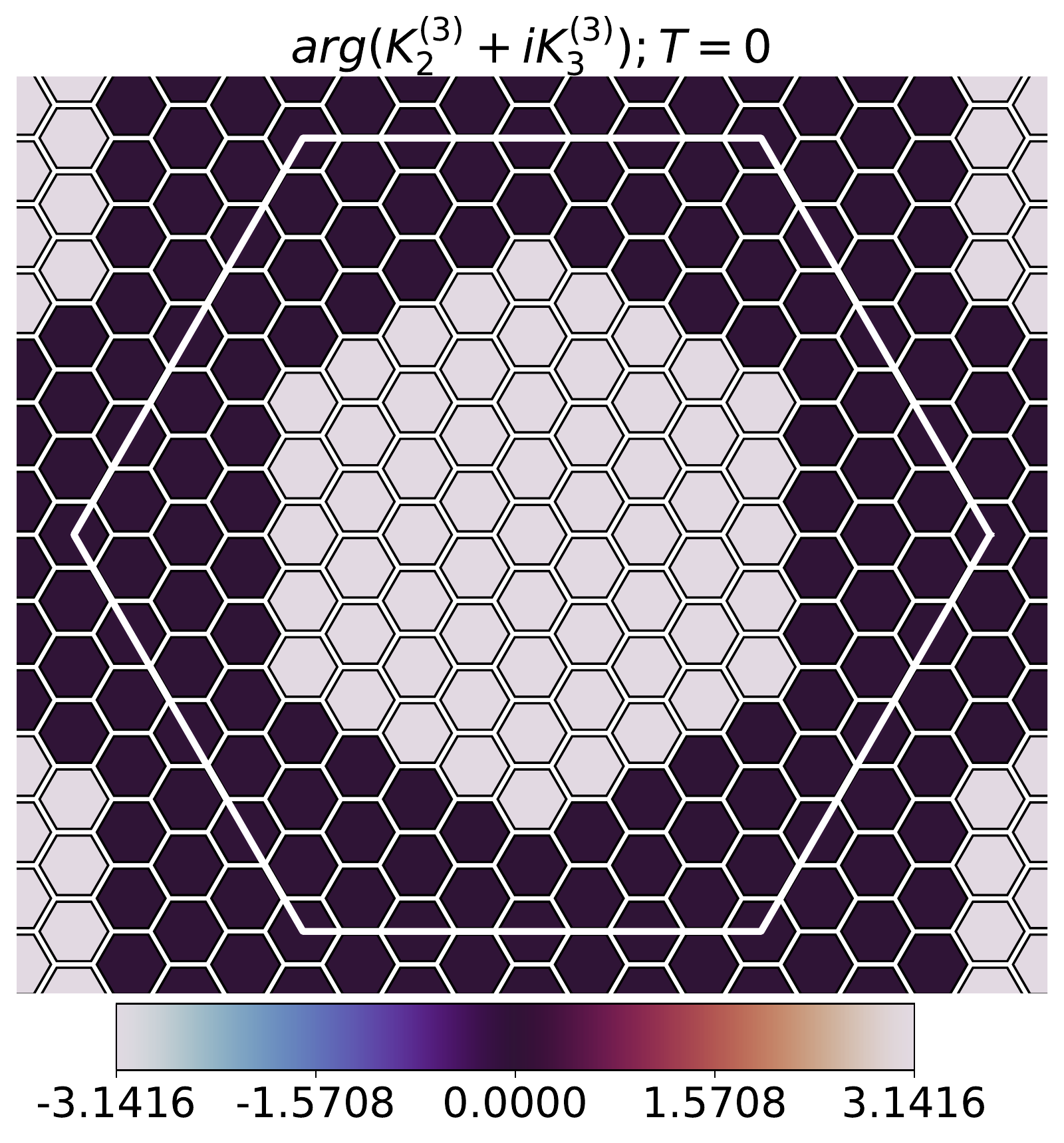}
}
\caption{\label{fig:KernelsTRS} Plots of the Kernels in Eq.~\ref{eq:KernelTRS} at $T=0$.}
\end{figure}

We numerically evaluate the kernel $\underline{K}^{(3)}(\kbs)$ in a $12\times 12$ momentum grid. The results are shown in Fig.~\ref{fig:KernelsTRS}. The largest eigenvalue is at $\kbs=0$ with eigenvector $[0, 1]^{\top}$. This corresponds to a valley-ferromagnetic state with $\ee^{3\ii \phi_i} = \pm \ii$. 

\newpage 

In the absence of $C_3$ symmetry of the ground state of $h_0$, $\Amc$ in general would be non-zero. The potential for $P_{CM}$ is now minimized by a $P_{CM}^* \propto \eps_3 \abs{\Phi_{3;\zero}}$ and $\q_{CM} = \Q_{3;\zero} - \arg(\Amc)$. The value of the potential at this minima is negative and proportional to $\eps_3^2 \abs{\Phi_{3;\zero}}^2$. This favours having a large $\abs{\Phi_{3;\zero}}$, {i.e.} a valley ferromagnetic order as does $e_2^{\triangle}$ when there is $C_3$ rotation symmetry. Therefore, we hope that even in the absence of $C_3$ symmetry, there could be a valley ferromagnet that breaks inversion symmetry and time-reversal symmetry. A main new feature in the absence of $C_3$, is that average crystal momentum $(P_{CM})$ is shifted, {i.e.} the band minima is displaced from its position at small $\rs$. 

\subsection{Berry curvature}\label{app:BerrCurvature}

Consider the Hamiltonian in Eq.~\ref{eq:EffBerryCurvature} of the main text. We treat $h_1$ and $h_2$ as perturbations to $h_0$. The leading order correction to the energy arises at first order in $\bar{\Bmc}_0$ and corresponds to $\bar{\Bmc}_0\expval{h_1}$, where $\expval{h_1}$ is the expectation value of $h_1$ on the ground state of $h_0$. This can be computed as 
\begin{equation}
    \begin{split}
        \expval{h_{1}} 
        &= -\sum_{ij} \frac{\expval{\mu_i(\eps P)_{i}K_{ij}Q_j+ \mu_jQ_jK_{ij}(\eps P)_{i}}}{2}\\  
        &= \frac{\ii}{4}\sum_{ij;aba'}
        \left( 
        \mu_i\eps_{aa'}\d_{a'b}\d_{ij}(K_{ij})_{ab}
        -  \mu_j\d_{ba'}\d_{ij}(K_{ij})_{ba}\eps_{aa'}\right)\\
        &= \frac{\ii}{4} \sum_{i; a} \left(\mu_i \eps_{ab}(K_{ii})_{ab}-\mu_j \eps_{ab}(K_{ii})_{ba}\right) = 0.
    \end{split}
\end{equation}

\subsection{Moir\'{e} corrections}
If we modify the dispersion relation to $\w \rightarrow \sqrt{\kMo+\w^2}$ and assume that $\kMo$ is large compared to the phonon bandwidth, we can perform a perturbative expansion in $\kMo$ for the kernel. In particular, 
\begin{widetext}
\begin{equation}
  \begin{split}
        \frac{\prod_{j=1,2,3}\sqrt{1+\eps x_j}}{\sum_{j=1,2,3}\sqrt{1+\eps x_j}} &= \frac{1}{3} +  \frac{\epsilon}{9} \left(x_1+x_2+x_3\right) 
  -\frac{5 \epsilon ^2}{108} \left(x_1^2-x_2 x_1-x_3 x_1+x_2^2+x_3^2-x_2 x_3\right)
    \\
   &
  +\frac{\left(34 (x_1^3+x_2^3+x_3^3)-21 (x_2 x_1^2+ x_3 x_1^2+ x_2^2 x_1+ x_3^2 x_1+ x_2 x_3^2 +x_2^2 x_3)+24  x_1 x_2 x_3\right) \epsilon
   ^3}{1296} + \OO(\eps^4)
  \end{split}
\end{equation}
\end{widetext}
Then we want to replace $\eps = 1/\kMo$ and identify the $x$'s with the $\w^2$'s and evaluate the averages. First, note that when the energy does not appear the average of $\cos(\cdots)$ and $\sin(\cdots)$ is zero because the two eigenvectors at the same $\qbs$-point have $\q_2-\q_1=\pi/2$. Secondly, the only terms that depend on $\kbs$ is the one that corresponds to that is the average of something proportional to $\w_{\a}^2\w_{\a'}^2\w_{\a''}^2$. In all the others, we can use the $\d_{\sum \qbs,\kbs}$ to get rid of the $\kbs$ dependence. So we can write $\underline{K}^{(3)}(\kbs) = K' \Id + \frac{1}{\kMo^2}\underline{\var{K}}^{(3)}(\kbs) +\OO(\kMo^{-3}) $ with
\begin{equation}
   \begin{split}
       \var{K}_{1}^{(3)}(\kbs)&= \frac{1}{\Nel^2}\sum_{\a} \d_{\kbs,\sum_{\a}\qbs_{\a}}   \frac{\w_{\a}^2\w_{\a'}^2\w_{\a''}^2}{54} \\
         \var{K}_{2}^{(3)}(\kbs)&= \frac{1}{\Nel^2}\sum_{\a} \d_{\kbs,\sum_{\a}\qbs_{\a}}   \frac{\w_{\a}^2\w_{\a'}^2\w_{\a''}^2}{54} \cos(2\sum_{\a}\q_{\s})\\
          \var{K}_{3}^{(3)}(\kbs)&= \frac{1}{\Nel^2}\sum_{\a} \d_{\kbs,\sum_{\a}\qbs_{\a}}   \frac{\w_{\a}^2\w_{\a'}^2\w_{\a''}^2}{54} \sin(2\sum_{\a}\q_{\s})
   \end{split}
\end{equation}
we again evaluate $\underline{\var{K}}^{(3)}(\kbs)$ numerically and find similar behaviour to Fig.~\ref{fig:KernelsTRS}. In particular, the eigenvector with the largest eigenvalue is the same as for $\underline{{K}}^{(3)}(\kbs)$.

\subsection{Extension to 3D}\label{app:WCin3d}

The above calculation easily generalizes to 3d. In particular, we consider electrons with inverse mass tensor $W$. Then define $\mst = 3\left(\Tr[W]\right)^{-1}$ and $W_i\mst= \Id + \lambda Q_i$ with $Q_i$ a traceless symmetric matrix. 

We obtain the same expression for the free energy and energy as in Sec.~\ref{app:MapToClockModel} with the following modifications. $A \in \{1,2,3\}$, the sum is over the 3 dimensional BZ.
\begin{equation}
    \var E=-\frac{\lambda^2}{32}\sum_{\kbs} {\Phi}_{\kbs}^{\dag} \underline{Y}(\kbs) {\Phi}_{\kbs}
\end{equation}
where ${\Phi}_{\kbs}$ is a five component vector with components $\Tr[{\tilde{D}}_{\kbs} S^a]$, where $\{S^a,a=1,2,3,4,5\}$ is a orthonormal basis for 3 by 3 symmetric traceless matrices. The kernel is 
\begin{equation}
    \underline{Y}(\kbs)_{ab}=\frac{1}{\Nel}\sum_{\kbs,\a,\a'} \d_{\kbs,\qbs+\qbs'} \frac{\w_{\a}\w_{\a'}}{\w_{\a}+\w_{\a'}}\Tr[\Pi_{\a}S^a\Pi_{\a'}(S^b)^{\dag}]
\end{equation}
In here $\w_{\a}$ and $\Pi_{\a}$ are the eigenvalues and projectors of the matrix $K(\kbs)$. We used the expressions of Refs.~\cite{PhysRev.122.1437,PhysRev.99.1128} to evaluate the kernel. When we evaluate the sum, we should not use the value of $\omega_{\alpha}$ when $\alpha = (\zero,A)$ but instead use an average around $\qbs =\zero$ because $K(\qbs)$ is discontinuous at $\qbs=0$ due to the long-range nature of Coulomb interactions. We approximate $K(\qbs \to \zero)_{ab} = K_0\frac{q_aq_b}{q^2}$ so that only the longitudinal model contributes. Then taking the average over all directions result in a zero contribution. Therefore, we simply omit the terms with $\qbs=\zero$ or $\qbs'=\zero$.

We approximate the kernel by an average over a $12\times12\times 12$ grid of the first BZ of the FCC crystal. We find that the largest eigenvalue of $\underline{Y}(\kbs)$ is at $\kbs=\Nbs =(\pi,\pi,0)$ (and 5 other symmetry related momenta) with eigenvalue, eigenvector pair 
\[
\Lambda \approx 0.4106 , V = \frac{1}{2}\left(
\begin{array}{ccc}
 0 & 0 & 1 \\
 0 & 0 & -1 \\
 1 & -1 & 0 \\
\end{array}
\right).
\]
At $\Gamma$, the diagonal matrices have eigenvalue $0.3773$ while the off-diagonal ones have $0.4054$. 

The kernel in the main text is $\tilde{J}(\kbs)= \frac{1}{32}\underline{Y}(\kbs)$.

\section{ Dispersion relations}\label{app:DispersionRelations}
In this section we give the details of calculation of small wavelength dispersion relation of the phonons. 
\def\z{q}
We focus on the frequency matrix around $\qbf=0$ for the case of period 2 stripes.
\begin{align}
    \Omega_{\qbf}^2 &= \Vmc \cdot \Kmc(\qbf)\cdot \Vmc \\
    \Vmc^2 &= \Wmc =  \frac{1}{2}\left(
\begin{array}{cc}
 W_A+W_B & W_A-W_B \\
 W_A-W_B & W_A+W_B \\
\end{array}
\right) \\
\Kmc(\qbs)&= \left(
\begin{array}{cc}
 K(\qbs) & 0 \\
 0 & K(\qbs+\Mbs) \\
\end{array}
\right) 
\end{align}
At $\qbs=0$, there are two zero eigenvalues. To leading order in $q$, $K(\qbs) = \z_0 \z\hat{\z}_a\hat{\z}_b$ and $K(\Mbs+\qbs)=K(\Mbs)$. We then expect that the order $\z$ perturbation will lift the degeneracy. First order degenerate perturbation theory tell us to compute the matrix elements of the perturbation in the basis of the degenerate states. In this case, we can take the basis as $\ebf_1=\Vmc^{-1}( 1,0,0,0)$ and $\ebf_2=\Vmc^{-1}( 0,1,0,0)$. Define $\tilde{M}$ as
\[
\tilde{M}_{ij} = (\Wmc^{-1})_{ij}; \quad i,j \in \{1,2\}.
\]
so that $\ebf_i\cdot\ebf_j=\tilde{M}_{ij}$. Let's assume this matrix is positive definite. Then we need to find the eigenvalues of
\[
\tilde{\W}^2_{\qbs}=
\frac{\z_0}{\z}\frac{1}{\sqrt{\tilde{M}}} \cdot \left(
\begin{array}{cc}
 \z_x\z_x & \z_x\z_y \\
 \z_y\z_x & \z_y\z_y \\
\end{array}
\right) \cdot \frac{1}{\sqrt{\tilde{M}}}.
\]
but as $\det(\tilde{\W}^2_{\qbs})=0$, there is a zero eigenvalue and the other one is given by the trace:
\begin{equation}
    \begin{split}
        \tilde{\w}_{\qbs,L}^{2} &= \frac{\z_0}{\z} \z\cdot\frac{1}{\tilde{M}} \cdot \z \\
    \tilde{\w}_{\qbs,T}^{2} &= 0.
    \end{split}
\end{equation}
Also note that because $\det(\W^2_{\qbs}) \propto q^3$ for $q$ small, we can find the leading order behaviour of $\w^2_{\qbs,T}$ to be $q^2$. To see this note that as a function of $\qbs$, the frequency of the optical phonons is 
\[
\w_{\qbs,op,i}^2 = \w_{0,op,i}^2 +\Omc(q^\a)\quad; i\in \{1,2\}, \w_{0,op,i}^2\neq 0.
\]
Then we calculate the determinant of $\W^2_{\qbs}$ in two different ways
\[
\w_{\qbs,L}^2\w_{\qbs,T}^2\w_{\qbs,op,1}^2\w_{\qbs,op,2}^2=\det(\W^2_{\qbs}) = \det(\Kmc(\qbs))\det(\Wmc),
\]
then expanding to leading order in $q$
\[
\w_{\qbs,T}^2 = \frac{\det(K(\qbs))}{\w^2_{\qbs,L}} \frac{\det(K(\Mbs))\det(\Wmc)}{\w^2_{0,op,1}\w^2_{0,op,2}}
\]
From here we can see that if $\tilde{M}$ is isotropic, the dispersion relation of both acoustic phonons is isotropic for $q \ll 1$.

For a symmetric matrix 
\[
 J = \left(
\begin{array}{cc}
 A & B \\
 B & A \\
\end{array}
\right) \Rightarrow
J^{-1} = \left(
\begin{array}{cc}
 (A-BA^{-1}B)^{-1} & \star \\
\star & (A-BA^{-1}B)^{-1} \\
\end{array}
\right)
\]
assuming $A$ and $B$ are invertible matrices.

If we identify $A = \Wmc_{\G}=\frac{1}{2}\left(W_{A}+W_B\right)$ and $B = \Wmc_{M}=\frac{1}{2}\left(W_{A}-W_B\right)$ we have 
\[
\tilde{M}^{-1}= W_{\G} - W_{M} W_{\G}^{-1} W_{M}.
\]

In particular, for the square lattice $W_\G = \tau^0$ and $W_M = \pm \l \t^3$ so $\tilde{M}^{-1} = (1-\l^2)\t^0 $. We can calculate the other frequencies by noting that $\Wmc = \id + \lambda  \t^3\mu^1 = \frac{2\h^{-\t^3\m^1}}{\h+1/\h}$ and $\Kmc(0) =\frac{(1+\mu^3)}{2} (A+B\t^3)$. Then
\[
\W_{0}^2 = \frac{2\h}{1+\h^2}\frac{(\eta^{\t^3 \mu^1}+\mu^3)}{2} (A+B\t^3) \Rightarrow \w_{0;i}^2 = k^{M_2}_{ii},
\]
in other words the optical frequencies of the phonons are independent of $\eta$.





\section{Optical conductivity}\label{app:OpticalConductivity}

In this Appendix, we simplify the expression for the optical conductivity that is used in Sec.~\ref{sec:nv_2} and Sec.~\ref{sec:nv_3}. First, note that 
\begin{align}
    J^a(\qbs)/e &= \sum_{i;b} (W_i)_{ab} \cdot p_i^b \ee^{\ii \qbs\cdot \Rbf_i} \\
    &=\sum_{i;b}\sum_{\Lbs\in \Lmc} \tilde{W}(-\Lbs)_{ab}\ee^{\ii (\Lbs+\qbs)\cdot \Rbf_i}p_i^b \\
    &= \sqrt{\Nel} \sum_{\Lbs \in \Lmc;b} \tilde{W}(-\Lbs)_{ab}P_{\Lbs+\qbs;b}\sqrt{\Enot \mst}\\
    &= \sqrt{\frac{\Enot}{\mst}}\sqrt{\Nel} \left[\Wmc \Pmc_{\qbs} \right]_{(a,0)}
\end{align}
Then the calculation reduces to calculate the matrix elements $\mel{m}{\Pmc_{\kbs}}{0}$ between the ground state $\ket{0}$ and the excited states $\ket{m}$. At every momenta $\kbs\in BZ'$ the Hamiltonian in units of $\Enot$ is 
\[
{\mathrm{H}_{\kbs}}/\Enot=\frac{1}{2}\Pmc_{\kbs}^\dag \Wmc \Pmc_{\kbs} + \frac{1}{2} \Qmc_{\kbs}^{\dag}\Kmc(\kbs) \Qmc_{\kbs}.
\]
Next we diagonalize the frequency matrix as $\W^2_{\kbs}=\Omc_{\kbs}^{\dag} D_{\kbs}^2\Omc_{\kbs}^{}$ with $\Omc_{\kbs}$ orthogonal and $D_{\kbs}$ diagonal. Then, we make the canonical change of variables $\acute{\Pmc}_{\kbs}=\Omc_{\kbs}\Vmc\Pmc_{\kbs}$ and $\acute{\Qmc}_{\kbs}=\Omc_{\kbs}(\Vmc)^{-1} \Qmc_{\kbs}$ to rewrite 
\[
{\mathrm{H}_{\kbs}}/\Enot=
\frac{1}{2}\acute\Pmc_{\kbs}^\dag \acute\Pmc_{\kbs} + \frac{1}{2} \acute\Qmc_{\kbs}^{\dag}D_{\kbs}^2\acute\Qmc_{\kbs}.
\]
Thus, 
\[
\mel{0}{(\acute\Pmc_{\kbs})_{\beta}}{n} = \sum_{\a} \d_{\omega_n,\Enot D_{\kbs;\a\a}} \ii\sqrt{\frac{ D_{\kbs;\a\a}}{2}}\d_{\a\b}
\]
which further implies
\[
\mel{0}{(\acute\Pmc_{\kbs})_{\beta}}{n} \overline{\mel{0}{(\acute\Pmc_{\kbs})_{\gamma}}{n} }=\sum_{\a} \d_{\omega_{n},\Enot D_{\kbs;\a\a}} {\frac{ D_{\kbs;\a\a}}{2}}\d_{\a\b}\d_{\a\g}.
\]
Then we multiply by $\d(\w-\w_n)/\w$ and sum over $n$ 
\begin{widetext}
\begin{align}
    \sum_{n}\frac{\d(\w-\w_n)}{\w}\mel{0}{(\acute\Pmc_{\kbs})_{\beta}}{n} \overline{\mel{0}{(\acute\Pmc_{\kbs})_{\gamma}}{n} }&=\sum_{\a,n} \frac{\d(\w-\w_{n})}{\w} \d_{\w_n,\Enot D_{\kbs;\a\a}} {\frac{ D_{\kbs;\a\a}}{2}}\d_{\a\b}\d_{\a\g} \notag\\
    &=\sum_{\a,n} \frac{\d(\w- \Enot D_{\kbs;\a\a})}{\Enot D_{\kbs;\a\a}}\d_{\w_n,\Enot D_{\kbs;\a\a}}  {\frac{ D_{\kbs;\a\a}}{2}}\d_{\a\b}\d_{\a\g} \notag\\
    &= \frac{1}{2 \Enot^2}\sum_{\a} \d(\w/\Enot-D_{\kbs,\a\a}) \d_{\b\a}\d_{\gamma\a} 
    \notag\\&=\frac{1}{2\Enot^2} \d(\w/\Enot - D_{\kbs})_{\b\gamma},
\end{align}
where we defined $\delta(\omega- X) \equiv \sum_{x} \delta(\omega-x)\Pi_{x}$ where $X$ is a Hermitian matrix, $x$ are the eigenvalues of $X$ and $\Pi_{x}$ are projections on the subspace with eigenvalue $x$.

The last step is to conjugate the above equality by $(\Omc \Vmc)^{-1}$ to go from $\acute\Pmc$ to $\Pmc$:
\begin{align}
    \sum_{n}\frac{\d(\w-\w_n)}{\w}\mel{0}{(\Pmc_{\kbs})_{\beta}}{n} \overline{\mel{0}{(\Pmc_{\kbs})_{\gamma}}{n} }
    &=\frac{1}{2\Enot^2} \left[\frac{1}{\Vmc}\Omc_{\kbs}^{\dag}\d(\w/\Enot - D_{\kbs}) \Omc_{\kbs}\frac{1}{\Vmc} \right]_{\b\gamma}\notag\\
    &= \frac{1}{2\Enot^2}\left[ \frac{1}{\Vmc} \d(\w/\Enot- \W_{\kbs}) \frac{1}{\Vmc}\right]_{\beta\gamma}
\end{align}
Finally, 
\begin{align}
    \s_{ab}(\w,\qbs) 
    &= \frac{4\pi}{A} \sum_{n} \frac{\d(\w-\w_n)}{\w} \mel{0}{J_a(\qbs)}{n}\overline{\mel{0}{J_b(\qbs)}{n}} \notag \\
    &= \frac{4\pi e^2}{ A}\frac{ \Nel }{2 \mst \Enot}\left[ \Wmc\frac{1}{\Vmc} \d(\w/\Enot- \W_{\qbs}) \frac{1}{\Vmc}\Wmc\right]_{(a,0),(b,0)} \notag\\
    &= \left(\frac{2\pi e^2 \nel}{\mst}\right)\frac{1}{\Enot}\left[ \Vmc \d(\w/\Enot- \W_{\qbs})\Vmc\right]_{(a,0),(b,0)}
\end{align}
where we used that $\nel=\Nel/A$.
\end{widetext}

\end{document}